\documentclass[a4paper,12pt]{article}
\usepackage{soul}
\usepackage[usenames,dvipsnames]{color}
\usepackage{jheppub}
\usepackage{amsmath, amssymb, slashed, epsf, color, graphicx, latexsym}
\usepackage{epsfig}
\usepackage{graphics}

\begin{document}
	
	\title{Large $D$ membrane for Higher Derivative Gravity and Black Hole Second Law}
		\author[a]{Yogesh Dandekar, }
	\author[b]{Arunabha Saha}
	\affiliation[a]{International Centre for Theoretical Sciences (ICTS-TIFR),
		Shivakote, Hesaraghatta Hobli,
		Bengaluru 560089, India} 
	\affiliation[b]{University of Geneva, 24 quai Ernest-Ansermet, 1211 Geneve 4, Switzerland}
	\emailAdd{yogesh.dandekar@icts.res.in}
	\emailAdd{arunabha.saha@unige.ch}

	\abstract{We derive the effective equations of the membranes dual to black holes in a particular theory of higher derivative gravity namely Einstein-Gauss-Bonnet (EGB) gravity at sub-leading order in $1/D$ upto linear order in the Gauss-Bonnet (GB) parameter $\beta$. We find an expression for an entropy current which satisfies a local version of second law onshell in this regime. We also derive the membrane equations upto leading order in $1/D$ but non-perturbatively in $\beta$ for EGB gravity. In this regime we write down an expression for a world-volume stress tensor of the membrane and also work out the effective membrane equation for stationary black holes.}
		

\maketitle
\section{Introduction}
For two derivative theory of gravity (Einstein-Hilbert theory) the area of the event horizon of a black hole monotonically increases (classically) throughout a physical process. Together with the fact that the ``first law" of black hole mechanics gives the entropy of a black hole to be proportional to the area of the horizon, gives the precise statement for the status of the second law of black hole thermodynamics for Einstein-Hilbert gravity: i.e. entropy always increases in a physical process. 

For higher derivative theories of gravity there is a candidate for entropy of black hole - namely the Wald entropy, which satisfies the first law of black hole mechanics \cite{Wald:1993nt, Iyer:1994ys}. But there is no general understanding of the status of the second law of black hole thermodynamics in these theories. Some candidates for entropy of black holes which satisfy the second law of black hole mechanics for specific type of dynamics involving black holes are known. e.g. in \cite{Bhattacharjee:2015yaa, Wall:2015raa}, an expression for entropy of black hole satisfying second law for small amplitude metric perturbations about stationary black hole configurations upto linear order in the amplitude were obtained. These expressions contain corrections over the Wald entropy which match with Dong-Camps terms \cite{Dong:2013qoa, Camps:2013zua} in holographic entanglement entropy. Again, in \cite{Bhattacharyya:2016xfs} a suitable candidate for second law was obtained upto an ``obstruction term". The obstruction term restricts the validity of the analysis to black hole dynamics for which the obstruction term is either negative or zero. e.g. The obstruction term is zero for dynamics which preserve spherical symmetry. The expressions for entropy obtained in both these situations reduce to the Wald entropy for stationary black holes and also reduce to the area of the event horizon in the limit of two derivative gravity.

On the other hand, during the last few years the large $D$ limit where one takes the number of spacetime dimensions $D\rightarrow\infty$ has been found to be particularly useful in studying dynamical processes involving black holes. In this limit the black hole dynamics can be described by a system of effective equations determining the dynamics of a finite number of variables of a non-gravitational system. There are two apparently different non-gravitational systems to which the black hole dynamics can be mapped to. In one the effective equations are obtained in terms of a dual object defined by its effective mass and momentum \cite{Emparan:2015gva, Suzuki:2015iha,Tanabe:2015isb,Emparan:2016sjk}\footnote{The effective equations in mass momentum formalism in $AdS$ space with boundary metric deformation has been worked out in \cite{Andrade:2018zeb}.}. The other formulation (which is the primary focus of this paper), recasts the black hole dynamics in terms of dynamics of a co-dimension one membrane propagating in the asymptotic spacetime of the black hole \cite{Bhattacharyya:2015dva, Bhattacharyya:2015fdk, Dandekar:2016fvw, Bhattacharyya:2017hpj, Bhattacharyya:2018szu,Kundu:2018dvx}. The membrane is characterised by its local shape and a world-volume velocity field. The effective equations in this case are a set of ``membrane equations" constraining the  membrane variables\footnote{In both the formalisms, we only study the late time dynamics of the black holes which correspond to time scales of the order of $\mathcal{O}(1/r_0)$ where, $r_0$ is the horizon size. The large $D$ limit confines the non-trivial part of the late time dynamics to a thin region of size $\mathcal{O}(r_0/D)$ about the horizon}. 

Interestingly, the effective membrane equations imply the second law of black hole thermodynamics for two derivative theories, without it being used as an input. It does even better in the sense that it gives the second law in terms of positive divergence of a local entropy current defined in terms of the membrane variables. The entropy current in this case is the obvious suspect: namely the generator of the horizon. The velocity field on the membrane world volume maps to the generator of the event horizon of the black hole and the membrane equation imply \cite{Dandekar:2016fvw}
\begin{eqnarray}
\hat{\nabla}\cdot u=\frac{2}{\mathcal{K}}\sigma_{\mu\nu}\sigma^{\mu\nu}+{\cal O}(1/D^2).
\end{eqnarray}
$\mathcal{K}$ is the trace of the local extrinsic curvature of the membrane embedded in the asymptotic spacetime of the black hole and $\sigma_{\mu\nu}$ is the shear tensor of the membrane velocity field $u^\mu$. $\hat{\nabla}$ denotes covariant derivative w.r.t the induced metric on the membrane. It can be shown \cite{Bhattacharyya:2016nhn} that in the large $D$ limit for uncharged black holes in asymptotic flat spacetime the local entropy current of the black hole to leading order in large $D$ is given by
\begin{equation}
J^\mu_{(s)}=\frac{u^\mu}{4}+{\cal O}(1/D^2)
\end{equation}
And hence, 
\begin{equation}
\hat{\nabla}\cdot J_{(s)}=\frac{1}{2\mathcal{K}}\sigma_{\mu\nu}\sigma^{\mu\nu}+\mathcal{O}(1/D^2)
\end{equation}
Hence, the membrane equation correctly predicts the positive definite local entropy production. 

The above procedure when applied to higher derivative theories of gravity can in principle lead us to a candidate for entropy which satisfies the second law of black hole thermodynamics for these theories\footnote{The more interesting possibility is if it can be shown that the membrane equations derived for a particular higher derivative theory of gravity does not allow for second law to be satisfied. In this case if we go by the principle that the second law cannot be violated in nature, this particular theory of higher derivative gravity should not be allowed in nature.}. 

The focus of this paper will be a particular theory of higher derivative gravity namely the Einstein-Gauss-Bonnet theory, the Lagrangian density for which is given by \footnote{We use indices $A,B,M,N$ to denote the coordinates over the spacetime. We use the $\mu,\nu,\alpha,\beta$ indices to denote the coordinates over the world-volume of the membrane.}
\begin{equation}
\mathcal{L}=\sqrt{-g}\left(R+\alpha \left(R^2+R_{ABCD}R^{ABCD}-4R_{AB}R^{AB}\right)\right).
\end{equation}
The leading order in $1/D$ membrane equation for this was worked out in \cite{Saha:2018elg} upto linear order in the Gauss-Bonnet parameter. It was shown there that for the solutions of the above theory to be continuously connected to the solutions of Einstein-Hilbert gravity in the $\alpha\rightarrow0$ limit $\alpha$ needs to be an $\mathcal{O}(1/D^2)$ quantity. A new variable $\beta$ was defined there such that 
\begin{equation}\label{asc}
\alpha=\frac{\beta}{(D-3)(D-4)}
\end{equation}
where, $\beta$ is an $\mathcal{O}(D^0)$ quantity \footnote{The leading order membrane equation for a general theory of four derivative gravity in presence of a cosmological constant was worked out in \cite{Kar:2019kyz}.}.The leading order in $1/D$ effective equation for EGB gravity non-perturbatively in the Gauss-Bonnet parameter in the mass momentum formalism were worked out in \cite{Chen:2017rxa}.

In this paper we work out the membrane equations to subleading order in $1/D$ upto linear order in $\beta$. These equations are given by

\begin{equation}\label{dotu}
\hat \nabla.u = \frac{2}{{\cal K}}\left(1-\beta \frac{{\cal K}^2}{D^2}\right)\sigma_{\alpha\beta}\sigma^{\alpha\beta} + 4 \beta \frac{{\cal K}}{D^2} \sigma_{\alpha\beta}K^{\alpha\beta} + 8\frac{\beta}{{\cal K}D^2} \left(\hat\nabla^\alpha\sigma_{\alpha\beta}\hat\nabla^\gamma{\sigma_\gamma}^\beta\right) + {\cal O}\left(\frac{1}{D^2}\right)
\end{equation}

\begin{equation}\label{vme}
	\begin{split}
	&\Bigg[\frac{\hat \nabla^2 u_\sigma}{{\cal K}} - \left(1-\beta\frac{{\cal K}^2}{(D-3)^2}\right)\frac{\hat\nabla_\sigma {\cal K}}{{\cal K}} + u^\alpha K_{\alpha\sigma} - \left(1+\beta\frac{{\cal K}^2}{(D-3)^2}\right) u\cdot \hat\nabla u_\sigma\Bigg]{\cal P}^\sigma_\gamma \\
	& + \Bigg[-\frac{u^\beta K_{\beta \delta} K^\delta_\sigma}{\cal{K}}+\left(1-\beta\frac{\mathcal{K}^2}{D^2}\right)\frac{\hat{\nabla}^2\hat{\nabla}^2 u_\sigma}{{\cal{K}}^3}-\frac{(\hat{\nabla}_\sigma{\cal{K}})(u\cdot\hat{\nabla}{\cal{K}})}{{\cal{K}}^3}-\frac{(\hat{\nabla}_\beta{\cal{K}})(\hat{\nabla}^\beta u_\sigma)}{{\cal{K}}^2}-\frac{2K^{\delta \sigma}\hat{\nabla}_\delta\hat{\nabla}_\sigma u_\sigma}{{\cal K}^2}\\
	&-\left(1+\beta\frac{\mathcal{K}^2}{D^2}\right)\frac{\hat{\nabla}_\sigma\hat{\nabla}^2{\cal{K}}}{{\cal{K}}^3}+\frac{\hat{\nabla}_\sigma(K_{\beta\delta} K^{\beta\delta} {\cal{K}})}{{\cal K}^3}+3\left(1+\beta\frac{5\mathcal{K}^2}{3D^2}\right)\frac{(u\cdot K\cdot u)(u\cdot\hat{\nabla} u_\sigma)}{{\cal{K}}}\\
	&-3\left(1+\beta\frac{\mathcal{K}^2}{D^2}\right)\frac{(u\cdot K\cdot u)(u^\beta K_{\beta\sigma})}{{\cal{K}}}-6\left(1+\beta\frac{7\mathcal{K}^2}{3D^2}\right)\frac{(u\cdot\hat{\nabla}{\cal{K}})(u\cdot\hat{\nabla} u_\sigma)}{{\cal{K}}^2}\\&+6\left(1+\beta \frac{5\mathcal{K}^2}{6D^2}\right)\frac{(u\cdot\hat{\nabla}{\cal{K}})(u^\beta K_{\beta\sigma})}{{\cal{K}}^2}+3\left(1+\beta\frac{8\mathcal{K}^2}{3D^2}\right)\frac{u\cdot\hat{\nabla} u_\sigma}{D-3}-3\left(1+\beta\frac{\mathcal{K}^2}{D^2}\right)\frac{u^\beta K_{\beta\sigma}}{D}\Bigg]{\cal P}^\sigma_\gamma \\
	& + \left(\beta\frac{ {\cal K}^2}{D^2}\right)\Bigg[ \frac{u^\alpha u^\beta \hat \nabla_\alpha \hat \nabla_\beta u_\sigma}{{\cal K}} + \frac{3}{{\cal K}}\frac{\hat{\nabla}^2 u_\beta}{\cal{K}}\sigma^\beta_\sigma+ \frac{7}{{\cal K}}\frac{\hat{\nabla}^2 u_\beta}{\cal{K}}\omega^\beta_\sigma+ \frac{u\cdot\hat{\nabla} u_\beta }{\cal{K}}\sigma^\beta_\sigma- 3\frac{u\cdot\hat{\nabla} u_\beta}{\cal{K}} \omega^\beta_\sigma   \\&- \frac{4}{\cal{K}}u\cdot\hat{\nabla} u_\beta {\cal P}^{\beta\alpha}K_{\alpha\sigma}  + \frac{3}{\cal{K}}u^\delta K_{\delta\beta}\sigma^\beta_\sigma + \frac{3}{\cal{K}}u^\delta K_{\delta\beta} \omega^\beta_\sigma+\left(-\frac{4}{D} + 9 \frac{u\cdot {\hat\nabla{\cal K}}}{{\cal K}^2} - 4 \frac{u\cdot K\cdot u}{{\cal K}}\right) \frac{\hat \nabla^2 u_\sigma}{\cal{K}}\Bigg]{\cal P}^\sigma_\gamma
	\\&  	= {\cal{O}}\left(\frac{1}{D}\right)^2
	\end{split}
	\end{equation} 
	where, $\omega_{\alpha\beta}$ is the vorticity of the membrane velocity field and $\mathcal{P}_{\mu\nu}$ is the projector orthogonal to the velocity field in the membrane world volume. 

Using the above membrane equation we find an entropy current given by 
\begin{eqnarray}\label{mec}
J_S^\mu = &&\frac{1}{4}\left(1+2\alpha \cal{R}\right)u^\mu - \alpha ~u_\nu \mathcal{R}^{\nu\mu} - \frac{2\alpha}{{\cal K}}\hat \nabla^\beta \sigma_{\beta\alpha}\sigma^{\alpha\mu} 
-\alpha ~u\cdot\hat \nabla u_\beta \hat\nabla^\beta u^\mu +{\cal O}\left(\frac{1}{D^3}\right)\nonumber\\
&&\text{where,}\nonumber\\ \mathcal{R}&&={\cal K}^2 - K_{\alpha\beta}K^{\alpha\beta}
\end{eqnarray}
which has a positive definite divergence given by \footnote{We will work in the large $D$ limit such that the dynamics preserves a large isometry and in this situation we find that the trace of the extrinsic curvature is positive definite. Also, working perturbatively in $\beta$, we have $\alpha\mathcal{R}\ll1$ and hence the RHS is positive definite.}
\begin{equation}\label{divec}
\hat{\nabla}\cdot J_S = \frac{1}{2{\cal K}} \left(1+\alpha {\cal R}\right)\sigma^2 + {\cal O}\left(\frac{1}{D^2}\right).
\end{equation}
Recall that $\alpha\sim\mathcal{O}(1/D^2)$ and hence the RHS above indeed starts at $\mathcal{O}(1/D)$. We will also show that the entropy corresponding to the above current in the stationary configuration is Wald entropy of the corresponding stationary black hole. From the positivity of divergence of the entropy current in the membrane world-volume, it can be shown that the integrated entropy always increases under membrane dynamics. Since, there is a duality between membrane and black hole dynamics, this quantity must also increase under black hole hole dynamics. Hence, the integrated charge obtained from this membrane entropy current is a candidate for black hole entropy satisfying second law for EGB gravity ( upto linear order in GB parameter). 

We also derive the membrane equation to leading order in $1/D$ and non-perturbatively in $\beta$ to be given by
\begin{equation}\label{npm}
\begin{split}
&\Bigg[\frac{\hat \nabla^2 u_\mu}{{\cal K}} - \left(1-\frac{\beta\frac{{\cal K}^2}{D^2}}{1+2\beta\frac{{\cal K}^2}{D^2}+2\beta^2\frac{{\cal K}^4}{D^4}}\right)\frac{\hat\nabla_\mu {\cal K}}{{\cal K}} + u^\alpha K_{\alpha\mu} \\&\quad\quad- \left(1+\frac{\beta\frac{{\cal K}^2}{D^2}}{1+2\beta\frac{{\cal K}^2}{D^2}+2\beta^2\frac{{\cal K}^4}{D^4}}\right) u\cdot \hat\nabla u_\mu\Bigg]{\cal P}^\mu_\sigma = {\cal O}\left(\frac{1}{D}\right) \\&\text{and,}\\& \hat \nabla\cdot u = {\cal O}\left(\frac{1}{D}\right)
\end{split}
\end{equation} 
Using this we derive an entropy current non-perturbative in $\beta$ given by
\begin{equation}
J^{\mu,NP}_S=\frac{u^\mu}{4}\left(1+2\alpha {\cal R}\right)-\alpha ~u_\nu {\cal R}^{\nu\mu} + {\cal O}(1/D^2)
\end{equation}
the divergence of which using the leading order membrane equation \eqref{npm} is  
\begin{equation}
\hat \nabla\cdot J^{NP}_S = {\cal O}\left(\frac{1}{D}\right)
\end{equation}

\section{The Einstein-Gauss-Bonnet Gravity}
Since, we will be working with black hole solutions of the Einstein-Gauss Bonnet (EGB) gravity, we will briefly review the them here. The action for the EGB gravity is given by
\begin{equation}
S=\int d^D x\sqrt{-g}\left(R+\frac{\beta}{(D-3)(D-4)} \left(R^2+R_{ABCD}R^{ABCD}-4R_{AB}R^{AB}\right)\right).
\end{equation}
This action is a member of the Lovelock-Lanczos family of theories of higher derivative gravity, all of which have the common property that their equation of motions has a maximum of two derivatives acting on the metric like Einstein-Hilbert theory. The equation of motion of EGB gravity is given by
\begin{eqnarray}\label{GB_eqn}
E_{AB}=&& R_{AB}-\frac{1}{2}g_{AB}R-\frac{\beta}{(D-3)(D-4)}\Bigg(\frac{1}{2}g_{AB}\left(R_{EFCD}R^{EFCD}-4R_{EF}R^{EF}+R^2\right)\nonumber\\
&&-2R R_{AB}+4R_{AC}R^C_B-4R^{CD}R_{CABD}-2R_{ACDF}R_B^{CDF}\Bigg)
\end{eqnarray}
The static spherically symmetric black hole solution for this theory in the Kerr-Schild coordinates is given by
\begin{eqnarray}\label{static_BH}
ds^2&=& -dt^2+dr^2+r^2d\Omega_{D-2}^2+(1-f(r))(dt+dr)^2\nonumber\\
f(r)&=&1+\frac{r^2}{2\beta}\left(1-\sqrt{1+\frac{4\beta r_h^{D-3}}{r^{D-1}}\left(1+\frac{\beta}{r_h^2}\right)}\right)
\end{eqnarray}
where $r_h$ is the radius of the horizon\footnote{It is easy to check that $$f(r_h)=0\quad \text{for}\quad \frac{\beta}{r_h^2}\ge-\frac{1}{2}$$ and $$f(r_h)=2+\frac{r_h^2}{\beta}\quad \text{for}\quad \frac{\beta}{r_h^2}<-\frac{1}{2}.$$Hence we require that $\beta\ge-\frac{r_h^2}{2}$ }. The Gauss-Bonnet (GB) parameter $\beta$ is $\mathcal{O}(D^0)$ as mentioned earlier. The necessity of this has been explained in \cite{Saha:2018elg} to make sure that the large $D$ solutions are smoothly connected to the solutions of Einstein-Hilbert theory in the $\beta\rightarrow 0$ limit. The basic reason being that the typical curvature squared objects are $\mathcal{O}(D^2)$ times the monomials in curvature term in the large $D$ limit. Hence, if $\beta$ has a leading order behaviour in the large $D$ limit which makes it grow faster than $\mathcal{O}(D^0)$ the GB terms will dominate over the Einstein-Hilbert term and the solutions will not be smoothly connected to the solutions of the Einstein-Hilbert theory  \footnote{Note that any other scaling where $\beta\sim \mathcal{O}(D^{-x})$, where, $x$ is a positive real number will also satisfy this criteria. So, in effect we are working with the strongest possible GB coefficient $\beta$ in the large $D$ limit which satisfies this criteria.}.
\section{The large $D$ dynamics of Black Holes in EGB gravity}
	In this section we briefly review the large $D$ membrane paradigm of \cite{Bhattacharyya:2015dva, Bhattacharyya:2015fdk, Dandekar:2016fvw, Bhattacharyya:2017hpj, Bhattacharyya:2018szu} aligned to the discussion in the context of higher derivative theories \cite{Saha:2018elg, Kar:2019kyz}. 
	We take the large $D$ limit by restricting the dynamics to finite number of directions so that the spacetime preserves a $SO(D-p-3)$ isometry as $D\rightarrow\infty$. Here, $p$ is an arbitrary order one positive integer.
	
	Before proceeding we would like to mention some notational conventions that we use. We use the indices $A,B,\ldots,M,N\ldots$ to denote the coordinates over the $D$ dimensional spacetime, which can be either black hole spacetime, or the background flat spacetime. $\nabla$ is the covariant derivative w.r.t the background flat spacetime. 
	We use the indices $\alpha,\beta\ldots,\mu,\nu,\ldots$ to denote the coordinates over the $D-1$ dimensional membrane world volume, which is embedded in the background flat spacetime . $g_{\mu\nu}$ denotes the induced metric on the membrane. We use the notation $\hat\nabla$ to denote the covariant derivative constructed from the induced metric on the membrane world-volume . ${\cal R}_{\mu\nu\lambda\sigma}$ denotes the Riemann curvature calculated from $g_{\mu\nu}$.
\subsection{Ansatz Metric}
The aim of the large $D$ membrane paradigm is to find dynamical solutions of the EGB equations in a perturbative manner in $1/D$. To do so  we need to choose a good starting ansatz metric on which to apply the perturbative expansion in $1/D$. Covariantisation of the Kerr-Schild form of the static black hole is a convenient ansatz metric. The ansatz metric for EGB gravity to linear order in $\beta$ is given by\footnote{This matches with the ansatz metric written in \cite{Saha:2018elg} as to leading order in $1/D$, $d\psi\cdot d\psi=\frac{\mathcal{K}^2}{D^2}$. Also, we require $\beta\frac{\mathcal{K}^2}{D^2}\ge-\frac{1}{2}$ for the horizon to be at $\psi=1$.}
	\begin{equation}\label{ansatz_metric}
ds^2=ds_{flat}^2+\left(\frac{(1+\beta~ d\psi\cdot d\psi)}{\psi^{D-3}}-\beta \frac{ d\psi\cdot d\psi}{\psi^{2(D-2)}}\right)\left(O_M dx^M\right)^2
\end{equation}
Here, $O_M=n_M-u_M$. $n_M$ is the unit normal\footnote{All normalisations, dot products and covariant derivatives in this section are w.r.t asymptotic flat spacetime of the black hole} to the surface $\psi=1$ which is horizon of the corresponding black hole and $u_M$ is a unit normalised time like vector orthogonal to $n_M$. $\mathcal{K}$ is the trace of the extrinsic curvature of the surface $\psi=1$ embedded in flat spacetime. In the coordinates of the metric \eqref{static_BH} 
\begin{eqnarray}
\psi=\frac{r}{r_h}, \quad n_M dx^M=dr, \quad u_M dx^M=-dt,\quad \mathcal{K}=\frac{D-2}{r_h}
\end{eqnarray}
As has been noted in \cite{Emparan:2014aba},  in the large $D$ limit all non-trivial physics is confined in a thin region of width $\frac{r_h}{D}$ outside the horizon of the black hole. From the ansatz metric we see that in the region where $\psi\gg 1$ the metric reduces to flat spacetime. But if we confine ourselves to regions where $\psi-1\sim\mathcal{O}(1/D)$ then $\psi^{-D}\sim e^{-D(\psi-1)}$. The ansatz metric reaches its asymptotic form exponentially fast along the normal to the horizon and has a non-trivial warping only upto a region of width $\mathcal{O}(1/D)$ outside the horizon. 

Since, we want to study dynamical processes involving black holes, we promote $\psi$ and $u$ to be functions of the spacetime points with the following restrictions
\begin{itemize}
	\item The derivatives of the functions $\psi$ and $u$ w.r.t the spacetime coordinates is $\mathcal{O}(D^0)$ and
	\item The metric ansatz preserves an $SO(D-p-3)$ isometry with $p$ held fixed at an arbitrary finite number as $D\rightarrow\infty$.
\end{itemize}
 The ansatz metric solves for the EGB equation to leading order in large $D$ \cite{Dandekar:2016fvw, Bhattacharyya:2017hpj, Saha:2018elg} provided
\begin{eqnarray}
 \nabla^2{\psi^{-(D-3)}}|_{\psi=1}=0 \quad \text{and}\quad \nabla\cdot u|_{\psi=1}=0
\end{eqnarray}
where the dot product and covariant derivatives are w.r.t the flat spacetime. 
The leading order ansatz metric can be thought of as being constructed by patching together locally boosted Schwarzschild solutions. The above two conditions are satisfied by the Schwarzschild black hole and hence to leading order they also need to be satisfied by our ansatz as to leading order they locally look like Schwarzschild metric. These conditions can be thought of as local constraints on the `shape' $\psi$ and `velocity' $u$ functions of a dual membrane in terms of which the black hole metric can be written. To leading order there is a duality between the black hole and a membrane provided the membrane satisfies the above two conditions. 

\subsection{The first order in $1/D$ corrections to the metric}
The ansatz metric \eqref{ansatz_metric} solves for the EGB equation only upto leading order in $1/D$ and not beyond that. Hence, we need correct the ansatz metric  so that it solves for the EGB equations upto the subleading order in $1/D$. The large $D$ regime makes it particularly simple to determine these metric corrections. 

The ansatz metric reaches its asymptotic value just outside the `membrane region'. Since, flat space solves for the EGB equation, we only need to add corrections to the metric which have non-zero support in the `membrane region'. We describe in brief the mechanism for this below. For the details of the procedure, one can look at the previous papers on the large $D$ membrane paradigm \cite{Dandekar:2016fvw, Bhattacharyya:2017hpj, Bhattacharyya:2018szu}\footnote{In practice, it is convenient to work in terms of a set of effective gravity equations in a reduced $p+3$ dimensional spacetime involving the $p+3$ dimensional metric and a dilaton field $\phi$.The results obtained this way can then be recast in terms of quantities in the full $D$ dimensional spacetime. The results obtained are then shown to be independent of the choice of $p$.The details of this procedure can be found in \cite{Bhattacharyya:2015dva, Bhattacharyya:2015fdk, Saha:2018elg, Kar:2019kyz}}. 

We use a coordinate system $(y^a)$ based around any point $x_0$ in the membrane region. 
\begin{equation}
x^M=x_0^M+\frac{\alpha_a^M y^a}{D}
\end{equation}
The coordinate system $y^a$ spans the entire patch region of length scale $\sim\mathcal{O}(1/D)$ in the membrane region about the point $x^0$ in the sense that as $y^a\rightarrow\infty$ we reach the boundary of the membrane region. 

The shape $(\psi)$ and the velocity function $(u^M)$ can be expressed in the region around the point $x_0$ as a Taylor series expansion given by
\begin{eqnarray}
&&\psi=1+\frac{y^M}{D}\partial_M\psi|_{x=x_0}+\frac{y^M y^N}{2D^2}\partial_M\partial_N\psi|_{x=x_0}+\frac{y^M y^N y^L}{6D^3}\partial_M\partial_N\partial_L \psi|_{x=x_0}+\ldots\nonumber\\
&&u^A=u_0^A|_{x=x_0}+\frac{y^M}{D}\partial_M u^A|_{x=x_0}+\frac{y^M y^N}{2D^2}\partial_M\partial_N u^A|_{x=x_0}+\frac{y^M y^N y^L}{6D^3}\partial_M\partial_N\partial_L u^A|_{x=x_0}+\ldots\nonumber\\
&& \text{where,}\quad y^M=\alpha^M_a y^a
\end{eqnarray}
In a similar manner the normal vector to the surface $\psi=1$ is given by 
\begin{equation}
n^A=n_0^A|_{x=x_0}+\frac{y^M}{D}\partial_M n^A|_{x=x_0}+\frac{y^M y^N}{2D^2}\partial_M\partial_N n^A|_{x=x_0}+\frac{y^M y^N y^L}{6D^3}\partial_M\partial_N\partial_L n^A|_{x=x_0}+\ldots
\end{equation}
where, the Taylor expansion coefficients of $n^A$ are appropriately related to the coefficients in the expansion of $\psi$. 
The Taylor expansion coefficients of $u^A$ are appropriately constrained to make the velocity vector unit normalised and orthogonal to the normal vector. 

Since,  the spacetime under consideration has a distinct fast direction along $d\psi$ we chose one of the patch coordinates $y^a$ along this direction. We call this coordinate $R$ and it is related to the shape function by
\begin{equation}
\psi=1+\frac{R}{D}.
\end{equation}
With this choice of the $R$ coordinate the blackening factor becomes
\begin{equation}
f=1-\psi^{-(D-3)}=1-e^{-R}+\mathcal{O}(1/D)
\end{equation}
The shape and velocity functions which had $\mathcal{O}(D^0)$ derivatives w.r.t the global coordinate $x^M$ have $\mathcal{O}(1/D)$ derivatives w.r.t the patch coordinates. Hence, it is easy to see that at first subleading order in $1/D$ the EGB equations when evaluated in the patch coordinates on the ansatz metric, evaluate to a non-zero quantity of the form
$$E^0_{MN}(\partial_M n_N|_{x=x_0},\partial_M u_N|_{x=x_0},R)+\beta E^1_{MN}(\partial_M n_M|_{x=x_0},\partial_M u_N|_{x=x_0},R)$$
\footnote{Although the derivatives of the shape and velocity fields are $\mathcal{O}(D^0)$, because of the large isometry in the $D\rightarrow\infty$ limit, the divergence of vector quantities can be $\mathcal{O}(D)$ higher than their naive order in $D$. See \cite{Dandekar:2016fvw} for more details.}Since, the part which remains unsolved at the subleading order is only a non-trivial function of the patch radial coordinate $R$ \footnote{though it contains ultra-local information about the local derivatives of the shape and velocity function at $x^0$.}, the necessary corrections need only be functions of the $R$ coordinates. i.e. the corrected metric at first subleading order can be written as
\begin{equation}
g_{MN}^{(1)}=g_{MN}^{ansatz}+\frac{1}{D}(H^0_{MN}(R)+\beta H^1_{MN}(R))
\end{equation}
We will be solving for the metric corrections order by order in $\beta$. It is easy to see that the corrected metric solves for the EGB equations at the subleading order in $1/D$ (to linear order in $\beta$) if the metric corrections satisfy the following ordinary differential equations(ODE) in the $R$ coordinates with sources and the various coefficients dependent on the ultra-local Taylor expansion shape and velocity data at the point $x_0$. The schematic form of these ODEs can be written as
\begin{eqnarray}\label{ODEfirst}
&&\mathcal{D}H^0_{MN}+E^0_{MN}(\partial_M n_N|_{x=x_0},\partial_M u_N|_{x=x_0},R)=0\nonumber\\
&&\mathcal{D}H^1_{MN}+E^1_{MN}(\partial_M n_N|_{x=x_0},\partial_M u_N|_{x=x_0},R)=0
\end{eqnarray}
where, $\mathcal{D}$ is an ordinary differential operator order two. The structure of the differential operator which is the same at all orders in $\beta$ is completely determined by the Einstein-Hilbert part of the EGB equations\footnote{If the computation is done in a non-perturbative manner in $\beta$, one would expect that the differential operator may contain more than two derivatives coming from the higher derivative terms in EGB equation.~But since, Gauss-Bonnet gravity falls into the Lovelock-Lanczos type theories, the maximum order of derivatives will be two even non-perturbatively. We will use this later to present result for first order membrane equations non-perturbatively in $\beta$.}.
\subsection{Some technical details}
\subsubsection{Choice of gauge}
We have to fix the gauge freedom in the choice of the metric corrections arising from the diffeomorphism invariance of gravity. We follow the convention followed in earlier papers and make the following gauge choice
\begin{equation}
H^0_{MN}O^M=0\quad H^1_{MN}O^M=0
\end{equation}
With this gauge the most general form of the metric correction can be written as
\begin{eqnarray}\label{metcorr}
&&H_{MN}=H^{(S)} O_M O_N+\frac{1}{D}H^{(Tr)}\mathcal{P}_{MN}+\left(H^{(V)}_M O_N+H^{(V)}_N O_M\right)+H^{(T)}_{MN}\nonumber\\
&& \text{where,}\quad \mathcal{P}_{MN}=\eta_{MN}+u_M u_N-n_M n_N\nonumber\\
&& \text{and,}\quad H_M^{(V)}n^M=H_M^{(V)}u^M=H_{MN}^{(T)}u^M=H_{MN}^{(T)}n^M=H_{MN}^{(T)}\mathcal{P}^{MN}=0
\end{eqnarray}
In the above and the rest of this section we have suppressed the superscript corresponding to the order in $\beta$. 
\subsubsection{Boundary conditions}
As stated earlier the metric corrections should have non-zero support only in the membrane region and hence we should impose the following boundary condition at the end of the membrane region
\begin{equation}
\lim_{R\rightarrow\infty}H_{MN}(R)=0
\end{equation}
Also, we are working in a coordinate system in which the ansatz metric is well defined everywhere outside the singularity at the centre of the black hole. We would like the corrections to also satisfy this property and hence we demand that the metric corrections be regular everywhere in the membrane region\footnote{We will not be interested in the metric corrections deep inside the black hole beyond the membrane region. Since, this region is causally disconnected from the observer outside the horizon we can describe physics outside the horizon without the knowledge of the metric corrections in the interior region.}. 

Even after imposing the above regularity and boundary conditions some ambiguity in the metric corrections is still left unfixed. This is due to the fact that the definitions of the velocity and shape field are ambiguous at subleading order in $1/D$ as these redefinitions leave the leading order ansatz metric unchanged. At leading order the velocity vector field raised w.r.t the flat spacetime is the generator of the event horizon of the spacetime located at $\psi=1$ \cite{Dandekar:2016fvw}. We fix the subleading ambiguity by demanding that the subleading order corrections keep the horizon at $\psi=1$\footnote{We demand that the surface $\psi=1$ remains null.} and keep the generator of the event horizon to be the velocity vector field. These two conditions impose the following two boundary conditions on the metric corrections at $R=0$ (or $\psi=1$). 
\begin{eqnarray}
&&H^{(S)}|_{R=0}=0\nonumber\\
&&H^{(V)}_M|_{R=0}=0.
\end{eqnarray}
After this the metric corrections are completely determined in terms of the shape and velocity function of the dual membrane modulo one remaining issue which we address below. 
\subsubsection{The auxiliary conditions}
The metric corrections are determined as functions of the $R$ coordinate with coefficients being the Taylor expansion coefficients of the membrane shape and velocity data in the patch coordinates about any arbitrary point in the membrane region. The Taylor expansion coefficients are local derivative data of the membrane shape and velocity vectors. These derivatives can also be along the normal direction itself. Now a membrane is solely determined by the data on the membrane hypersurface only. Derivatives of the membrane data orthogonal to the hypersurface are not well-defined. The only way to make sense of the normal derivatives of the membrane data is to uplift their definition to a family of hypersurfaces, of which the membrane is just one member. We call the choice of this uplift as the auxiliary conditions. Obviously this uplift is not unique and hence different choices will give rise to different values for the normal derivatives and hence apparently different metric corrections. But the `physical content' of the apparently different metrics are the same as they all arise from the same physical membrane data.


In this paper we choose to work with the auxiliary conditions used in \cite{ Bhattacharyya:2017hpj} given by
\begin{eqnarray}
&&\nabla^2\left(\frac{1}{\psi^{(D-3)}}\right)=0\nonumber\\
&& O\cdot \nabla O\cdot\mathcal{P}=0	
\end{eqnarray}
where, $\mathcal{P}$ is the projector orthogonal to $u$ and $n$ mentioned above. This choice was found to be particularly useful in the sense that for Einstein gravity, the metric corrections at first subleading order vanished and the next to the subleading order corrections were also very simple. Once, this choice has been made the metric of the black hole is now completely defined in terms of the data of the dual membrane. 
\section{The results at first order in $1/D$}
The large $D$ membrane effective equations to leading order and the first subleading corrections to the metric of black holes in EGB gravity (upto linear order in $\beta$) was obtained in \cite{Saha:2018elg}\footnote{Look for corresponding results for the most general theory of four derivative gravity with cosmological constant in \cite{Kar:2019kyz}.}.  The procedure is based on classifying the gravity equations into dynamical and constraint equations along the ``direction of evolution'', which in our case is in the direction of coordinate $R$ (which also happens to be the direction of normal to the membrane). The interesting property of the constraint equations is that they have homogeneous parts with order one even though the rest of the dynamical EGB equations are order two. Hence, the constraint equations are thought of as a set of constraint on the possible ``initial" conditions on the metric on a hypersurface rather than as evolution equations. Another interesting property is that once the constraint equations are solved on a hypersurface and the dynamical equations are solved everywhere, the constraint equations are naturally solved on all other hypersurfaces orthogonal to evolution. Hence, the information contained in the constraint equations is hypersurface invariant. Following \cite{Dandekar:2016fvw} we evaluate the constraint equations on the $R=0$ hypersurface, where, the homogeneous parts vanish, and so one is left with a set of constraint on the membrane data given by\footnote{The equation obtained using this procedure from the constraint EGB equations are in terms of the full spacetime derivatives. Those equations have to be pulled back to the membrane world volume using the auxiliary conditions mentioned above. Only after this the membrane equations become well defined in the sense that they track the world-volume dynamics of the membrane variables. The equation presented here is this pulled back form.}
\begin{eqnarray}\label{first_order_memeq}
&&\Bigg[\frac{\hat \nabla^2 u_\mu}{{\cal K}} - \left(1-\beta\frac{{\cal K}^2}{D^2}\right)\frac{\hat\nabla_\mu {\cal K}}{{\cal K}} + u^\alpha K_{\alpha\mu} - \left(1+\beta\frac{{\cal K}^2}{D^2}\right) u\cdot \hat\nabla u_\mu\Bigg]{\cal P}^\mu_\sigma = {\cal O}\left(\frac{1}{D}\right)\nonumber\\
&&\text{and,}\nonumber\\
&&  \hat\nabla\cdot u=\mathcal{O}(1/D)
\end{eqnarray}
The corrected metric with the choice of auxiliary conditions used in this paper are given below \footnote{In \cite{Saha:2018elg} a different set of auxiliary conditions were used: $1) n\cdot \nabla n_M=0$ and $2)n\cdot \nabla u_M=0$. And hence the metric corrections obtained there were also different.} (upto linear order in $\beta$) by 
\begin{equation}\label{first order metric}
\begin{split}
g_{MN} &= \eta_{MN} + \left(\frac{(1+\beta d\psi\cdot d\psi)}{\psi^{D-3}}-\beta \frac{ d\psi\cdot d\psi}{\psi^{2(D-2)}}\right)\left(O_M dx^M\right)^2\\&
 -4\beta e^{-2R}\left(-1+e^{R}\right)\frac{{\cal K}}{(D-3)^2}\left(\frac{u\cdot \nabla {\cal K}}{{\cal K}}-u\cdot K\cdot u+\frac{{\cal K}}{4(D-3)}\right)O_MO_N 
\\& + \beta R e^{-R}\frac{{\cal K}}{(D-3)^2} \left(\frac{\nabla^2 u_A}{{\cal K}}-2 u\cdot \nabla u_A + u^BK_{BA} \right)\left({\cal P}^A_MO_N+{\cal P}^A_NO_M\right)
\\& -2 \beta e^{-R}\frac{{\cal K}}{(D-3)^2}\left(K_{AB}-\nabla_{(A}u_{B)} - \frac{K_{CD}-\nabla_{(C}u_{D)}}{D-2}{\cal P}^{CD}{\cal P}_{AB}\right){\cal P}^A_M{\cal P}^B_N
\end{split}
\end{equation}
Note that in the $\beta\rightarrow0$ limit the subleading correction is zero which is consistent with the fact that in two derivative gravity the first subleading corrections to the metric are zero with the choice of auxiliary conditions used here \cite{Bhattacharyya:2017hpj}. 
\section{The results at second order in $1/D$}
In \cite{Dandekar:2016fvw} the algorithm to systematically correct the metric upto arbitrary orders in $1/D$ was presented for Einstein-Hilbert gravity. This algorithm can be easily adapted to the EGB equations. This algorithm involves using the corrected metric at a given order as the seed metric at next order. In our case we add corrections to the first order corrected metric \eqref{first order metric} in the same form as \eqref{metcorr} but at one higher order in $1/D$. The structure of the homogeneous parts of the ODEs on the corrections remain the same at all orders \cite{Dandekar:2016fvw}, but the sources now depend on higher Taylor expansion coefficients of the membrane and velocity data. The schematic form of the differential equations at this order ( like in \eqref{ODEfirst}) are given by
\begin{eqnarray}\label{ODEsecond}
&&\mathcal{D}H^0_{MN}+E^0_{MN}(\partial_M n_N|_{x=x_0},\partial_M u_N|_{x=x_0},\partial_M\partial_N n_L|_{x=x_0},\partial_M\partial_N u_L|_{x=x_0},R)=0\nonumber\\
&&\mathcal{D}H^1_{MN}+E^1_{MN}(\partial_M n_N|_{x=x_0},\partial_M u_N|_{x=x_0},\partial_M\partial_N n_L|_{x=x_0},\partial_M\partial_N u_L|_{x=x_0},R)=0\nonumber\\
\end{eqnarray}

\subsection{The membrane equations at first subleading order}
One can again evaluate the constraint part of the above equations at $R=0$ to obtain the  subleading correction to the membrane equation of motion. These are given by
\begin{eqnarray}\label{sme}
\hat \nabla.u = \frac{2}{{\cal K}}\left(1-\beta \frac{{\cal K}^2}{D^2}\right)\sigma_{\alpha\beta}\sigma^{\alpha\beta} + 4 \beta \frac{{\cal K}}{D^2} \sigma_{\alpha\beta}K^{\alpha\beta} + 8\frac{\beta}{{\cal K}D^2} \left(\hat\nabla^\alpha\sigma_{\alpha\beta}\hat\nabla^\gamma{\sigma_\gamma}^\beta\right) + {\cal O}\left(\frac{1}{D^2}\right)\nonumber\\
\end{eqnarray}
The vector membrane equation is given by
\begin{equation}\label{vme}
	\begin{split}
		&\Bigg[\frac{\hat \nabla^2 u_\sigma}{{\cal K}} - \left(1-\beta\frac{{\cal K}^2}{(D-3)^2}\right)\frac{\hat\nabla_\sigma {\cal K}}{{\cal K}} + u^\alpha K_{\alpha\sigma} - \left(1+\beta\frac{{\cal K}^2}{(D-3)^2}\right) u\cdot \hat\nabla u_\sigma\Bigg]{\cal P}^\sigma_\gamma \\
		& + \Bigg[-\frac{u^\beta K_{\beta \delta} K^\delta_\sigma}{\cal{K}}+\left(1-\beta\frac{\mathcal{K}^2}{D^2}\right)\frac{\hat{\nabla}^2\hat{\nabla}^2 u_\sigma}{{\cal{K}}^3}-\frac{(\hat{\nabla}_\sigma{\cal{K}})(u\cdot\hat{\nabla}{\cal{K}})}{{\cal{K}}^3}-\frac{(\hat{\nabla}_\beta{\cal{K}})(\hat{\nabla}^\beta u_\sigma)}{{\cal{K}}^2}-\frac{2K^{\delta \sigma}\hat{\nabla}_\delta\hat{\nabla}_\sigma u_\sigma}{{\cal K}^2}\\
		&-\left(1+\beta\frac{\mathcal{K}^2}{D^2}\right)\frac{\hat{\nabla}_\sigma\hat{\nabla}^2{\cal{K}}}{{\cal{K}}^3}+\frac{\hat{\nabla}_\sigma(K_{\beta\delta} K^{\beta\delta} {\cal{K}})}{{\cal K}^3}+3\left(1+\beta\frac{5\mathcal{K}^2}{3D^2}\right)\frac{(u\cdot K\cdot u)(u\cdot\hat{\nabla} u_\sigma)}{{\cal{K}}}\\
		&-3\left(1+\beta\frac{\mathcal{K}^2}{D^2}\right)\frac{(u\cdot K\cdot u)(u^\beta K_{\beta\sigma})}{{\cal{K}}}-6\left(1+\beta\frac{7\mathcal{K}^2}{3D^2}\right)\frac{(u\cdot\hat{\nabla}{\cal{K}})(u\cdot\hat{\nabla} u_\sigma)}{{\cal{K}}^2}\\&+6\left(1+\beta \frac{5\mathcal{K}^2}{6D^2}\right)\frac{(u\cdot\hat{\nabla}{\cal{K}})(u^\beta K_{\beta\sigma})}{{\cal{K}}^2}+3\left(1+\beta\frac{8\mathcal{K}^2}{3D^2}\right)\frac{u\cdot\hat{\nabla} u_\sigma}{D-3}-3\left(1+\beta\frac{\mathcal{K}^2}{D^2}\right)\frac{u^\beta K_{\beta\sigma}}{D}\Bigg]{\cal P}^\sigma_\gamma \\
		& + \left(\beta\frac{ {\cal K}^2}{D^2}\right)\Bigg[ \frac{u^\alpha u^\beta \hat \nabla_\alpha \hat \nabla_\beta u_\sigma}{{\cal K}} + \frac{3}{{\cal K}}\frac{\hat{\nabla}^2 u_\beta}{\cal{K}}\sigma^\beta_\sigma+ \frac{7}{{\cal K}}\frac{\hat{\nabla}^2 u_\beta}{\cal{K}}\omega^\beta_\sigma+ \frac{u\cdot\hat{\nabla} u_\beta }{\cal{K}}\sigma^\beta_\sigma- 3\frac{u\cdot\hat{\nabla} u_\beta}{\cal{K}} \omega^\beta_\sigma   \\&- \frac{4}{\cal{K}}u\cdot\hat{\nabla} u_\beta {\cal P}^{\beta\alpha}K_{\alpha\sigma}  + \frac{3}{\cal{K}}u^\delta K_{\delta\beta}\sigma^\beta_\sigma + \frac{3}{\cal{K}}u^\delta K_{\delta\beta} \omega^\beta_\sigma+\left(-\frac{4}{D} + 9 \frac{u\cdot {\hat\nabla{\cal K}}}{{\cal K}^2} - 4 \frac{u\cdot K\cdot u}{{\cal K}}\right) \frac{\hat \nabla^2 u_\sigma}{\cal{K}}\Bigg]{\cal P}^\sigma_\gamma
		\\&  	= {\cal{O}}\left(\frac{1}{D}\right)^2
	\end{split}
\end{equation} 
where, $$\sigma_{\alpha\beta}=\mathcal{P}^\mu_\alpha\mathcal{P}^\nu_\beta\frac{\hat{\nabla}_\mu u_\nu+\hat{\nabla}_\nu u_\mu}{2}-\frac{\mathcal{P}_{\alpha\beta}}{D-2}\hat{\nabla}\cdot u$$ is the shear and
$$\omega_{\alpha\beta}=\mathcal{P}^\mu_\alpha\mathcal{P}^\nu_\beta\frac{\hat{\nabla}_\mu u_\nu-\hat{\nabla}_\nu u_\mu}{2}$$ is the vorticity of the membrane velocity field $u^\mu$. The first line in the vector membrane equation is the leading order in $1/D$ contribution to the membrane equation and the rest are first subleading order in $1/D$ corrections to the membrane equation. 
 \subsection{The metric corrections}
 The metric corrections \eqref{metcorr} are written in terms of components with various tensor structures w.r.t the hypersurface orthogonal to the velocity vector $u$ and the normal vector $n$. Since, the differential equations for the corrections are Ordinary linear differential equations, the metric components in different tensor structures do not mix. The sources to the homogeneous parts of each sector also preserve the tensor structure. Below we present the differential equations obtained in each tensor structure. We write down the explicit form of the homogeneous parts of the equations here with a schematic form for the corresponding sources. We also present the solutions to these equations which preserve the right boundary and regulation conditions here in an integrated form in terms of the schematic sources like in \cite{Dandekar:2016fvw}. For the detailed formulas for the sources, the reader is referred to the Appendix \eqref{detailedsources}. 
\subsubsection{The tensor sector}
Let $\mathcal{E}_{MN}=0$ be the ordinary differential equations obtained by applying the EGB equations to the first subleading order in $1/D$ metric plus the second order metric corrections. Then the decoupled equation for the tensor sector is obtained from the following combination 
\begin{equation}
\mathcal{P}^A_M \mathcal{P}^B_N\mathcal{E}_{AB}-\frac{\mathcal{P}_{MN}}{D-2}\mathcal{P}^{AB}\mathcal{E}_{AB}
\end{equation}
 The decoupled tensor equations at different orders in $\beta$ are given by
 \begin{eqnarray}
 &&-\frac{{\cal K}^2}{2D^2}\left(\frac{dH^{(0,T)}_{MN}}{dR} + (1-e^{-R})\frac{d^2H^{(0,T)}_{MN}}{dR^2}\right)+\mathcal{S}_{MN}^{(0,T)}(R)=0\nonumber\\
 &&\text{and,}\nonumber\\
 &&-\frac{{\cal K}^2}{2D^2}\left(\frac{dH^{(1,T)}_{MN}}{dR} + (1-e^{-R})\frac{d^2H^{(1,T)}_{MN}}{dR^2}\right)+\mathcal{S}_{MN}^{(1,T)}(R)\nonumber\\
 &&\quad+\frac{\mathcal{K}^4e^{-2R}}{2D^4}\left(\frac{dH^{(0,T)}_{MN}}{dR}+(e^{R}-1)\frac{d^2H^{(0,T)}_{MN}}{dR^2}\right)=0\nonumber\\
 \end{eqnarray}
 where, the superscript over the metric corrections denote the order in powers of $\beta$ and the tensor structure respectively. $\mathcal{S}_{MN}^{(0,T)}(R)$ and $\mathcal{S}_{MN}^{(1,T)}(R)$ denote the corresponding sources for the differential equations in terms of membrane data. The equation at linear order in $\beta$ contains an extra source piece coming from the solution of the metric correction at zeroth order in $\beta$. Hence, the procedure involves first solving for the zeroth order in $\beta$ metric corrections and then using it to solve for the first order in $\beta$ metric corrections while imposing the regularity and boundary conditions at each order. 

In \cite{Dandekar:2016fvw} a detailed algorithm to obtain solutions to the differential equations appearing in two derivative theory of gravity in an integrated form for arbitrary sources with the boundary conditions mentioned above was presented. Since, the structure of the homogeneous part in our case is also completely determined by the two derivative part of the EGB equations, we can use the solutions presented there adapted to the sources for EGB gravity. We present the solution here.
\begin{equation}
H^{(0,T)}_{MN} = -2 \frac{D^2}{{\cal K}^2} \int_{R}^{\infty} \frac{dy}{e^y-1}\int_{0}^{y} dx~ e^x~ \mathcal{S}^{(0,T)}_{MN}(x)
\end{equation}
and
\begin{equation}
H^{(1,T)}_{MN} = -2 \frac{D^2}{{\cal K}^2} \int_{R}^{\infty} \frac{dy}{e^y-1}\int_{0}^{y} dx~ e^x~ \hat{\mathcal{S}}^{(1,T)}_{MN}(x)
\end{equation}
where, $$\hat{\mathcal{S}}^{(1,T)}_{MN}(R)=\mathcal{S}_{MN}^{(1,T)}(R)+\frac{\mathcal{K}^4}{2D^4}\left(e^{-2R}\frac{dH^{(0,T)}_{MN}}{dR}+(e^{-R}-e^{-2R})\frac{d^2H^{(0,T)}_{MN}}{dR^2}\right)$$
\subsubsection{The vector sector}
The decoupled equation for the vector sector is obtained from the combination $u^A \mathcal{E}_{AB} {\cal P}^B_M$ and is given by
\begin{eqnarray}
&&-\frac{{\cal K}^2}{2D^2}(1-e^{-R})e^{-R}\frac{d}{dR}\left(e^R\frac{dH_M^{(0,V)}}{dR}\right)+\mathcal{S}^{(0,V)}_M(R)=0\nonumber\\
&&-\frac{{\cal K}^2}{2D^2}(1-e^{-R})e^{-R}\frac{d}{dR}\left(e^R\frac{dH_M^{(1,V)}}{dR}\right)+\mathcal{S}^{(1,V)}_M(R)- \frac{\mathcal{K}^4}{2D^4}(1-e^{-R})\frac{d }{dR}\left(e^{-R}\frac{d H_M^{(0,V)}}{dR}\right)=0\nonumber\\
\end{eqnarray}
Using the results of \cite{Dandekar:2016fvw} the solutions to the above equations can be written as
\begin{equation}
H^{(0,V)}_M(R) = -2 \frac{D^2}{{\cal K}^2} \int_{R}^{\infty} dy~ e^{-y} \int_{0}^{y} dx~ \frac{e^{2x}}{e^x-1} \mathcal{S}^{(0,V)}_M + e^{-R} C^{(0,V)}_M 
\end{equation}
where, $C^{(0,V)}_M$ is chosen so that $H^{(0,V)}_M(R=0)=0$. The solution for the linear in $\beta $ part of metric corrections is
\begin{equation}
H^{(1,V)}_M = -2 \frac{D^2}{{\cal K}^2} \int_{R}^{\infty} dy~ e^{-y} \int_{0}^{y} dx~ \frac{e^{2x}}{e^x-1} \hat{\mathcal{S}}^{(1,V)}_M + e^{-R} C^{(1,V)}_M 
\end{equation}	
where, $$\hat{\mathcal{S}}^{(1,V)}_M(R)=\mathcal{S}^{(1,V)}_M(R)- \frac{\mathcal{K}^4}{2D^4}(1-e^{-R})\frac{d }{dR}\left(e^{-R}\frac{d H_M^{(0,V)}}{dR}\right)$$
and $C^{(1,V)}_M$ is chosen so that $H^{(1,V)}_M(R=0)=0$.  
 \subsubsection{The scalar sector}
 The scalar sector contains two independent metric correction components, namely: $H^{(Tr)}$ and $H^{(S)}$. 
 The decoupled equation for $H^{(Tr)}$ is obtained via the combination $O^M O^N\mathcal{E}_{MN}$ and is given by
  \begin{eqnarray}
&& -\frac{{\cal K}^2}{2D^2}\frac{d^2}{dR^2}H^{(0,Tr)} + \mathcal{S}^{(0,Tr)} = 0 \nonumber\\
&&-\frac{{\cal K}^2}{2D^2}\frac{d^2}{dR^2}H^{(1,Tr)} + \mathcal{S}^{(1,Tr)} -\frac{\mathcal{K}^4}{D^4}e^{-R}\frac{d^2}{dR^2}H^{(0,Tr)} = 0 \nonumber
 \end{eqnarray}
 The solution for the above equations can be written as 
 \begin{equation}
 H^{(0,Tr)} = 2 \frac{D^2}{{\cal K}^2} \int_{R}^{\infty} dy \int_{y}^{\infty} dx~ \mathcal{S}^{(0,Tr)}
 \end{equation}
 and
 \begin{equation}
 H^{(1,Tr)} = 2 \frac{D^2}{{\cal K}^2} \int_{R}^{\infty} dy \int_{y}^{\infty} dx~ \hat{\mathcal{S}}^{(1,Tr)}
 \end{equation}
 where,
 $$\hat{\mathcal{S}}^{(1,Tr)}(R)= \mathcal{S}^{(1,Tr)} -\frac{\mathcal{K}^4}{D^4}e^{-R}\frac{d^2}{dR^2}H^{(0,Tr)}$$
 There is no decoupled equation available for $H^{(S)}$. We use the combination of EGB equation given by $O^M u^N \mathcal{E}_{MN}$ to solve for the $H^{(S)}$ corrections once all other corrections are solved for\footnote{In \cite{Dandekar:2016fvw} the constraint equation along the $O$ vector was used to solve for $H^{(S)}$ }. The equation at order ${\cal O}(\beta^0)$ is given by 
 \begin{equation}
 \frac{{\cal K}^2}{2D^2}\left(\frac{d}{dR}H^{(0,S)}+\frac{d^2}{dR^2}H^{(0,S)}\right) + \hat{{\cal S}}^{(0,S)} = 0
 \end{equation}
 where, 
 \begin{equation}
\hat{{\cal S}}^{(0,S)} = -\frac{{\cal K}^2}{4D^2}e^{-R}\frac{d}{dR}H^{(Tr_0)}+\frac{{\cal K}^2}{2D^2}\frac{\nabla^A H^{(V_0)}_A}{\cal K} + \mathcal{S}^{(0,S)}
 \end{equation}
 The equation at order ${\cal O}(\beta^1)$ is given by
 \begin{equation}
 \frac{{\cal K}^2}{2D^2}\left(\frac{d}{dR}H^{(1,S)}+\frac{d^2}{dR^2}H^{(1,S)}\right) + \hat{{\cal S}}^{(1,S)} = 0
 \end{equation}
 where,
 \begin{equation}
 \begin{split}
 \hat{{\cal S}}^{(1,S)} &=  \frac{{\cal K}^4}{D^4}e^{-2R}\Bigg[-e^R\left(\frac{d}{dR}H^{(0,S)}-\frac{d^2}{dR^2}H^{(0,S)}\right)-\frac{\nabla^AH^{(0,V)}_A}{{\cal K}}+\left(1-\frac{e^R}{4}\right)\frac{d}{dR}H^{(0,Tr)}\\&+(-1+e^R)\frac{d^2}{dR^2}H^{(0,Tr)}\Bigg] -\frac{{\cal K}^2}{4D^2}e^{-R}\frac{d}{dR}H^{(1,Tr)}+\frac{{\cal K}^2}{2D^2}\frac{\nabla^A H^{(1,V)}_A}{\cal K} + \mathcal{S}^{(1,S)}
 \end{split}
 \end{equation}
 The solutions are given by
 \begin{equation}
 H^{(0,S)} = \frac{2D^2}{{\cal K}^2} \int_{R}^{\infty} dy~ e^{-y} \int_{0}^{y} dx~ e^x \hat{{\cal S}}^{(0,S)} + e^{-R} C^{(0,S)} 
 \end{equation}
 where, $C^{(0,S)}$ is chosen so that $H^{(0,S)}(R=0)=0$ and
 \begin{equation}
 H^{(1,S)} = \frac{2D^2}{{\cal K}^2} \int_{R}^{\infty} dy~ e^{-y} \int_{0}^{y} dx~ e^x \hat{{\cal S}}^{(1,S)} + e^{-R} C^{(1,S)} 
 \end{equation}
 where,  $C^{(1,S)}$ is chosen so that $H^{(1,S)}(R=0)=0$.
 \section{The Entropy Current}
 For the Einstein-Hilbert gravity it was shown in \cite{Bhattacharyya:2016nhn} that the membrane entropy current for the black hole in the dual membrane picture is proportional to the velocity vector field of the membrane. More precisely, it was shown that to leading order the membrane entropy current is given by
 \begin{eqnarray}
 J^\mu_S=\frac{u^\mu}{4}+\mathcal{O}(1/D^2)
 \end{eqnarray}
 And using subleading order membrane equations for two derivative gravity derived in \cite{Dandekar:2016fvw} it can be shown that
 \begin{equation}
 \hat\nabla\cdot J_S=\frac{1}{2\mathcal{K}}\sigma_{\mu\nu}\sigma^{\mu\nu}+\mathcal{O}(1/D^2)
 \end{equation}
 The above equation can be thought of as a local version of the second law of black hole thermodynamics. Hence, we get the second law out of the membrane equations without it being used as an explicit input. 
 
 Since, no candidate expression for second law ( in full generality) for EGB gravity is known, the task will be to write down the expression for an entropy current which has a non-trivial positive definite divergence in the large $D$ limit (upto linear order in $\beta$). This entropy current must satisfy some general criteria, namely
 \begin{itemize}
 	\item In the limit of $\beta\rightarrow0$ the entropy current should reduce to the corresponding quantity for Einstein-Hilbert gravity. 
 	\item In the limit of stationary ``dynamics" the integrated entropy from the membrane picture should match the Wald entropy of the corresponding stationary black hole.  
 \end{itemize}
To do this it will be convenient to recast the scalar membrane equation into a more useful form.  

Consider the membrane propagating in background flat spacetime. Let $\cal{R}_{\alpha\beta}$ be the intrinsic Ricci curvature of the membrane world-volume. According to the Gauss-Codacci equation (see e.g. Eq.(3.39) of \cite{poisson_2004}) we have
\begin{eqnarray}\label{mic}
\cal{R}_{\alpha\beta}& =& {\cal K}K_{\alpha\beta} - K_{\alpha\mu}{K^{\mu}}_{\beta}\nonumber\\
\implies{\cal R}&=&\mathcal{K}^2-K_{\alpha\beta}K^{\alpha\beta}\nonumber\\
\text{and,}\quad  u\cdot {\cal R}\cdot u&=&\mathcal{K}u\cdot K\cdot u-u\cdot K\cdot K\cdot u
\end{eqnarray} 
Using this we can recast the scalar membrane equation \eqref{sme} as
\begin{equation}\label{dotue}
\hat \nabla\cdot u = \frac{2}{{\cal K}}\left(1- \alpha {\cal R}\right)\sigma_{\alpha\beta}\sigma^{\alpha\beta} + 4 \alpha~ \sigma_{\alpha\beta}{\cal R}^{\alpha\beta} + \frac{8\alpha}{{\cal K}} \left(\hat{\nabla}^\alpha\sigma_{\alpha\beta}\hat{\nabla}^\gamma{\sigma_\gamma}^\beta\right) + {\cal O}\left(\frac{1}{D^2}\right)
\end{equation}
It is clear from the above equation that the entropy current is not the same as in Einstein-Hilbert gravity as the divergence of the velocity field is not manifestly positive definite for any membrane configuration. 

We propose that an entropy current for the membrane dual to black holes in EGB gravity upto linear order in $\beta$ is given by
	\begin{equation}
\begin{split}
J_S^\mu &= \frac{1}{4}\left(1+2\alpha \cal{R}\right)u^\mu - \alpha ~u_\alpha \mathcal{R}^{\alpha\mu} - \frac{2\alpha}{{\cal K}}\hat \nabla^\beta \sigma_{\beta\alpha}\sigma^{\alpha\mu} -\alpha ~u\cdot\hat \nabla u_\beta \hat\nabla^\beta u^\mu +{\cal O}\left(\frac{1}{D^3}\right)
\\& = \frac{1}{4}u^\mu - \alpha ~u_\alpha \left(\mathcal{R}^{\alpha\mu}-\frac{1}{2}\mathcal{R}g^{\alpha\mu}\right) - \frac{2\alpha}{{\cal K}}\hat \nabla^\beta \sigma_{\beta\alpha}\sigma^{\alpha\mu} -\alpha ~u\cdot\hat \nabla u_\beta \hat\nabla^\beta u^\mu +{\cal O}\left(\frac{1}{D^3}\right)
\end{split}
\end{equation}
The divergence of the above entropy current is given by
\begin{eqnarray}
\hat{\nabla}\cdot J_S&=&\frac{\hat{\nabla}\cdot u}{4}-\alpha\nabla_\mu u_\alpha\left(\mathcal{R}^{\alpha\mu}-\frac{1}{2}\mathcal{R} g^{\alpha\mu}\right)-\frac{2\alpha}{\mathcal{K}}(\hat{\nabla}\cdot \sigma)^2-\alpha ~u\cdot \hat{\nabla} u^\beta \mathcal{R}_{\alpha \beta}u^\alpha + {\cal O}(1/D^2)\nonumber\\
\end{eqnarray}
In the above we substitute the value of $\hat{\nabla}\cdot u$ from the scalar membrane equation \eqref{dotue} and use the following identity 
\begin{equation}
{\cal R}^{\mu\alpha}\hat{\nabla}_\mu u_\alpha=\sigma_{\alpha\beta}\mathcal{R}^{\alpha\beta}+u\cdot\hat{\nabla}u_\alpha \mathcal{R}^{\alpha\mu}u_\mu
\end{equation}
to get
\begin{equation}
\hat{\nabla}\cdot J_S=\frac{\sigma_{\alpha\beta}\sigma^{\alpha\beta}}{2\mathcal{K}}\left(1+\alpha{\cal R}\right)+{\cal O}(1/D^2).
\end{equation}
Hence, the proposed entropy current has a positive definite divergence\footnote{We confine ourselves to dynamics which preserve a large isometry in the $D\rightarrow\infty$ limit. In this limit the leading order contribution to the trace of the extrinsic curvature comes from the curvature of the large sphere in the isometry directions. Hence, the trace of the extrinsic curvature is always positive definite in the situations that we consider. Also, at leading order in $1/D$, ${\cal R}={\cal K}^2$ and hence positive definite.}.
The current written above reduces to the entropy current for the two derivative Einstein-Hilbert gravity in the limit where, the GB parameter $\beta\rightarrow 0$ (at leading order in $1/D$). Hence, it satisfies one of the consistency conditions. In the next subsection we demonstrate the second consistency condition, namely the match with the Wald entropy of the dual black holes in the stationary limit.

 \subsection{Comparison with Wald Entropy}
 We define the stationary configuration of the membrane as the one where there is no entropy production. In terms of the entropy current defined above these configurations would correspond to zero divergence of the entropy current. From the above section we see that the stationary configuration (like in the Einstein-Hilbert gravity) corresponds to the membrane configuration where the shear of the velocity field vanishes. We also have
 \begin{equation}
 \hat{\nabla}\cdot u=0
 \end{equation}
 Since, the velocity field is both shear and divergenceless, it must be proportional to a timelike Killing vector field present in the membrane world-volume \cite{Caldarelli:2008mv}. So, the existence of stationary configuration necessitates the presence of a timelike Killing vector field in the membrane world-volume. 
 
 Since, the entropy current is conserved in the stationary configuration we can compute the entropy which is the charge corresponding to this current on any arbitrary space-like slice of the membrane world-volume. Since, the membrane world-volume has a timelike Killing vector field, we can choose a `Kaluza-Klein' like coordinate system (used also in \cite{Dandekar:2017aiv}), in which the induced metric on the membrane world-volume is given by
 \begin{equation}
 ds^2_{ind}=-e^{2\sigma(x)}\left(dt+a_i(x)dx^i\right)^2+f_{ij}(x)dx^i dx^j
 \end{equation}
 It is clear from the form of this metric that the $\frac{\partial}{\partial t}$ is the Killing vector field. Let us now construct the entropy by integrating the entropy current over the  $t=$constant spacelike slice, $\Sigma_t$. Let $q_\mu$ be the unit normal to the slice $\Sigma_t$ in the membrane world-volume. 
 \begin{equation}
 S_{mem}=\int_{\Sigma_t}\sqrt{h}J^\mu q_\mu=\int_{\Sigma_t}\sqrt{f}e^{\sigma(x)}J^t.
 \end{equation}
 In the above we have used some results associated with the Kaluza-Klein type metric from section $6$ of \cite{Dandekar:2017aiv}. Using the fact that we are in the stationary configuration, the entropy for EGB gravity in the Kaluza-Klein type coordinates then evaluates to
 \begin{equation}
 S_{mem}=\int_{\Sigma_t}\sqrt{f}\left(\frac{1}{4}\left(1+2\alpha\mathcal{R}\right)-\alpha\mathcal{R}_t^t\right)
 \end{equation}
 Now, using the results in section $2.1$ of \cite{Banerjee:2012iz} we have \footnote{In this section the quantities with overhead bar are w.r.t. the metric $f_{ij}$.}
 \begin{eqnarray}\label{rdi}
 \mathcal{R}&=&\bar{\mathcal{R}}-2\bar{\nabla}^2\sigma+\mathcal{O}(1)
 \end{eqnarray}
 Also,
 \begin{eqnarray}
 \mathcal{R}_t^t&=&{\mathcal{R}^i}_{tit}g^{tt}+{\mathcal{R}^\mu}_{t\mu j}g^{t j}\nonumber\\
 \text{where,}\quad  {\mathcal{R}^i}_{tit}&=&\bar{\Gamma}^i_{ij}e^{2\sigma}f^{jm}\partial_m\sigma+\mathcal{O}(1)=-e^{2\sigma}\bar{\nabla}^2\sigma +\mathcal{O}(1)\nonumber\\
 \text{and}\quad {\mathcal{R}^\mu}_{t\mu j}&=& \bar{\Gamma}^k_{kl}e^{2\sigma}f^{lk}\partial_k\sigma a_j+\mathcal{O}(1)=-e^{2\sigma}\bar{\nabla}^2\sigma a_j+\mathcal{O}(1)
 \end{eqnarray}
Combining the expressions above we have,
\begin{equation}\label{rtt}
\mathcal{R}_t^t=(-e^{-2\sigma}+a^2)(-e^{2\sigma}\bar{\nabla}^2\sigma) -a^j(-e^{2\sigma}\bar{\nabla}^2\sigma a_j)=\bar{\nabla}^2\sigma + {\cal O}(1)
\end{equation}
 where, $\bar{\mathcal{R}}$ is the Ricci scalar of the metric $f_{ij}$. Hence, from \eqref{rdi} and \eqref{rtt} we have 
 \begin{equation}
 S_{mem}=\int_{\Sigma_t}\sqrt{f}\left(\frac{1}{4}\left(1+2\alpha\bar{\mathcal{R}}\right)+{\cal O}(1/D^2)\right)
  \end{equation}
We now turn our focus on the Wald entropy for stationary black holes in EGB gravity which is given by the following expression  (e.g.\cite{Bhattacharyya:2016xfs})
 \begin{equation}\label{waldbh}
 S_{Wald}=\frac{1}{4}\int_{\Sigma_v}\sqrt{\bf{h}}\left(1+2\alpha~\bf{\cal{R}}_v\right).
 \end{equation}
 Here, $\Sigma_v$ are spatial slices of the horizon at constant $v$ such that $\partial_v$ is the generator of the event horizon. $\bf{\mathcal{R}}_v$ is the intrinsic curvature of the induced metric on $\Sigma_v$ thought of as being embedded in the black hole spacetime.
 
 In our conventions, the spacetime metric on the horizon of the black hole is given by
\begin{equation}
g_{MN} = \eta_{MN} + O_M O_N + H^{(T)}_{MN}  + \frac{1}{D} H^{(Tr)}{\cal P}_{MN} 
\end{equation} 
the above metric is rank two as the event horizon is a null hypersurface. It was shown in \cite{Bhattacharyya:2016nhn} (section $7$) that for computing dot products of all vectors tangent to the event horizon the above metric is equivalent to the metric given by
\begin{equation}
\mathbf{H}_{MN} = {\cal P}_{MN} + H^{(T)}_{MN}  + \frac{1}{D} H^{(Tr)}{\cal P}_{MN} 
\end{equation} 
For all practical purposes the above metric can also be taken to be the metric on the slice $\Sigma_v$ mentioned above. From our computations we know that
\begin{equation}
H^{(Tr)}=H^{(0,Tr)}+\beta H^{(1,Tr)}\quad , \quad H^{(T)}_{MN}  = H^{(0,T)}_{MN}+ \beta H^{(1,T)}_{MN} 
\end{equation}
where,
\begin{equation}
H^{(0,Tr)}={\cal O}\left(\frac{1}{D^3}\right)~~ ,~~ H^{(1,Tr)}={\cal O}\left(\frac{1}{D^2}\right)~~,~~H^{(0,T)}_{MN}={\cal O}\left(\frac{1}{D^2}\right)~~,~~H^{(1,T)}_{MN}={\cal O}\left(\frac{1}{D}\right)
\end{equation} 
For a metric which can be written as
\begin{equation}
g_{MN}=\bar{g}_{MN}+k_{MN}+\mathcal{O}(k^2)
\end{equation}
the determinant is given by
\begin{equation}
\sqrt{g}=\sqrt{\bar{g}}\left(1+\frac{1}{2}\bar{g}^{MN}k_{MN}\right)
\end{equation}
and the Ricci Scalar is given by
\begin{equation}
{\mathcal{R}}_g=-\mathcal{R}^{MN}_{\bar{g}} k_{MN}+\bar{\nabla}^M\bar{\nabla}^N k_{MN}+\bar{\nabla}^2k^M_M
\end{equation}
Hence, upto $\mathcal{O}(1/D^2)$ and upto $\mathcal{O}(\beta)$ we get
\begin{eqnarray}
&&\sqrt{\bf{h}}=\sqrt{\mathcal{P}}|_{\Sigma_v}\left(1+\frac{\beta}{2}H^{(1,Tr)}\right)|_{\Sigma_v}\nonumber\\
&&\mathcal{R}_{\bf v}=\mathcal{R}_{\mathcal{P}}+\mathcal{O}(1) \quad (\because \text{trace and divergences add orders of $D$})
\end{eqnarray}
We must remember that due to the isometry of the configurations $\mathcal{R}_{\bf v}$ is $\mathcal{O}(D^2)$. Hence, the Wald entropy on the membrane world-volume is given by
\begin{equation}
S_{Wald}=\frac{1}{4}\int_{\Sigma_v}\sqrt{\mathcal{P}}\left(1+2\alpha\mathcal{R}_{\mathcal{P}}\right)
\end{equation}
Going back once again to the Kaluza-Klein type metric in the membrane picture the projector orthogonal to $u$ vector ($\partial_t$ vector in this case), is the metric $f_{ij}$. Hence, the membrane entropy can also be written in a more covariant form as
\begin{equation}
 S_{mem}=\frac{1}{4}\int_{\Sigma_t}\sqrt{\mathcal{P}}\left(1+2\alpha\mathcal{R}_\mathcal{P}\right)
\end{equation}
To leading order in $\beta$ the Wald entropy matches the entropy obtained from the membrane picture if $$\mathcal{P}|_{\Sigma_v}=\mathcal{P}|_{\Sigma_t}.$$
Once the above is assumed it naturally follows that the intrinsic curvature calculated will also be the same for these metrics and hence we have a match between Wald entropy and membrane entropy to linear order in $\beta$ upto subleading order in $1/D$. 


\section{Some results non-perturbative in $\beta$ for EGB gravity}
Until now all the results presented by us at second order in $1/D$ are valid upto linear order in $\beta$ in a perturbative expansion. We obtain a membrane entropy current which has a manifestly positive definite divergence upto $\mathcal{O}({\beta})$ 1at $\mathcal{O}(1/D)$. The natural question that comes to mind is if there will still exist an entropy current with positive definite divergence non-perturbatively in $\beta$. The first step in this direction is to find the membrane equation upto first subleading order in $1/D$ non-perturbatively in $\beta$. In this section we present the leading order membrane equations non-perturbatively in $\beta$ for EGB gravity. 

The starting point for the computation is the non-perturbative Kerr-Schild form of the Schwarzschild metric for EGB gravity given by
\begin{eqnarray}
ds^2&=& -dt^2+dr^2+r^2d\Omega_{D-2}^2+(1-f(r))(dt+dr)^2\nonumber\\
\text{where,}\quad f(r)&=&1+\frac{r^2}{2\beta}\left(1-\sqrt{1+\frac{4\beta r_h^{D-3}}{r^{D-1}}\left(1+\frac{\beta}{r_h^2}\right)}\right)
\end{eqnarray}
using this we write the ansatz metric  as
\begin{eqnarray}
ds^2_{ansatz}=ds^2_{flat}-\frac{D^2}{2\beta \mathcal{K}^2}\psi^2\left(1-\sqrt{1+4\beta\frac{\mathcal{K}^2}{D^2}\frac{1}{\psi^{D-1}}\left(1+\beta \frac{\mathcal{K}^2}{D^2}\right)}\right)(O_M dx^M)^2\nonumber.\\
\end{eqnarray}
Next we apply the algorithm mentioned in the earlier sections to correct the metric at sub-leading order. In this case we do not grade the metric corrections in a perturbative series in $\beta$. 
Due to the special property of Einstein-Gauss-Bonnet (EGB) gravity and more generally of Lovelock-Lanczos type gravity, the ordinary differential equations on the metric corrections are still of order two and the constraint equations are still order one. Both the coefficients of the homogeneous part of the ODEs and the sources are much more complicated in this case. The homogeneous parts of the relevant constraint equations vanish when evaluated at $R=0$ ( assuming that the metric corrections are regular everywhere) . We are again left with a set of constraint on the membrane data: namely the leading order in $1/D$ membrane equations non-perturbatively in $\beta$ given by
\begin{equation}\label{npme}
\begin{split}
&\Bigg[\frac{\hat \nabla^2 u_\mu}{{\cal K}} - \left(1-\frac{\beta\frac{{\cal K}^2}{D^2}}{1+2\beta\frac{{\cal K}^2}{D^2}+2\beta^2\frac{{\cal K}^4}{D^4}}\right)\frac{\hat\nabla_\mu {\cal K}}{{\cal K}} + u^\alpha K_{\alpha\mu} \\&- \left(1+\frac{\beta\frac{{\cal K}^2}{D^2}}{1+2\beta\frac{{\cal K}^2}{D^2}+2\beta^2\frac{{\cal K}^4}{D^4}}\right) u\cdot \hat\nabla u_\mu\Bigg]{\cal P}^\mu_\sigma = {\cal O}\left(\frac{1}{D}\right) \\&\text{and}\\& \hat \nabla\cdot u = {\cal O}\left(\frac{1}{D}\right)
\end{split}
\end{equation} 
Once the membrane equations are satisfied the  ODEs for the metric corrections can be solved with the same boundary and regularity conditions mentioned earlier for the perturbative case. We do not present the results for the metric corrections here as the solutions are very long and do not help in gaining any new understanding of the black hole dynamics relevant at this order in $1/D$. 
\subsection{Leading order entropy current at non-perturbative order}
Inspired by the form of the entropy current derived above at subleading order in $1/D$ (upto linear order in $\beta$), we can write down an entropy current at leading order in $1/D$ but non-perturbatively in $\beta$ given by
\begin{equation}
J^{\mu,NP}_S=\frac{u^\mu}{4}\left(1+2\alpha{\cal R}\right)-\alpha~u_\alpha \mathcal{R}^{\alpha\mu} + {\cal O}(1/D^2)
\end{equation}
 This current has the right $\beta\rightarrow 0$ limit in the sense that it matches the entropy current of the Einstein-Hilbert gravity. The divergence of the above entropy current using the non-perturbative membrane equations is given by
\begin{eqnarray}
\hat{\nabla}\cdot J^{NP}_S&=&\frac{\hat{\nabla}\cdot u}{4}-\alpha\hat{\nabla}_\mu u_\alpha\left(\mathcal{R}^{\alpha\mu}-\frac{1}{2}g^{\alpha\mu}\mathcal{R}\right)\nonumber\\
&=&\mathcal{O}(1/D)\quad (\because \hat{\nabla}\cdot u=\mathcal{O}(1/D))
\end{eqnarray}
i.e. there is no entropy production at leading order in large $D$ for EGB gravity. This is similar to result obtained for Einstein-Hilbert gravity. To understand if there exists an entropy current for EGB gravity which satisfies second law we need to evaluate the non-perturbative in $\beta$ membrane equation of motion upto subleading order in $1/D$. We leave this task for a future project. 

\subsection{A world-volume stress tensor}
A correct world volume stress tensor\footnote{A world-volume stress tensor is defined only in the world-volume of the membrane. It is not the stress tensor of the membrane in the spacetime and hence cannot be used to compute the gravitational radiation due to the membrane in the spacetime.} for the membrane at a given order in $1/D$ has the property that its conservation equations are satisfied on-shell.  From the form of the non-perturbative in $\beta$ membrane equation at leading large $D$ order presented above one can guess the form of the world-volume stress tensor to the relevant order as
\begin{equation}\label{mst}
\begin{split}
&16 \pi T_{\mu\nu} = (1+\alpha{\cal K}^2){\cal K} u_\mu u_\nu + \left(\frac{1-B}{1+B}\right)(1+\alpha{\cal K}^2) K_{\mu\nu} -2\left(\frac{1+\alpha{\cal K}^2}{1+B}\right) \sigma_{\mu\nu} \\&\quad\quad\quad\quad + A_\mu u_\nu + A_\nu u_\mu \\ & B=\frac{\alpha{\cal K}^2}{1+2\alpha{\cal K}^2+2\alpha^2{\cal K}^4} \quad , \quad A_\mu = -4 \alpha{\cal K}^2 \left(\frac{1+\alpha{\cal K}^2}{1+2\alpha{\cal K}^2}\right) u^\alpha K_{\alpha\beta}{\cal P}^\beta_\mu,\quad \alpha=\frac{\beta}{(D-3)(D-4)}
\end{split}
\end{equation} 
The conservation equation of this stress tensor along the velocity vector field is given by
\begin{eqnarray}
16\pi\hat\nabla^\mu T_{\mu\nu}u^\nu&=&-(1+\alpha \mathcal{K}^2) \mathcal{K} \hat\nabla\cdot u-u\cdot \hat\nabla\mathcal{K}(1+\alpha\mathcal{K}^2)-2 \alpha\mathcal{K}^2u\cdot\hat\nabla \mathcal{K}\nonumber\\
&&+\frac{1+\alpha\mathcal{K}^2+2\alpha \mathcal{K}^4}{1+2\alpha\mathcal{K}^2}u\cdot\hat\nabla\mathcal{K}-2\frac{1+2\alpha\mathcal{K}^2+2\alpha^2\mathcal{K}^4}{1+2\alpha \mathcal{K}^2}\hat\nabla^\mu\sigma_{\mu\nu}u^\nu
\end{eqnarray}
Also, $$\hat\nabla^\mu\sigma_{\mu\nu}u^\nu=\mathcal{O}(1).$$
Combining the above two we get
\begin{eqnarray}
16\pi\hat\nabla^\mu T_{\mu\nu}u^\nu&=&-(1+\alpha\mathcal{K}^2)\mathcal{K}\hat\nabla\cdot u+\mathcal{O}(1)\nonumber\\
&=&\mathcal{O}(1)\quad (\because \hat\nabla\cdot u=\mathcal{O}(1/D))
\end{eqnarray}
Similarly, the conservation equation orthogonal to the velocity vector field is given by
\begin{eqnarray}
16\pi\hat\nabla^\mu T_{\mu\nu}\mathcal{P}^\nu_\alpha&=&(1+\alpha\mathcal{K}^2)\mathcal{K} u\cdot \hat\nabla u_\nu \mathcal{P}^\nu_\alpha+\frac{1+\alpha\mathcal{K}^2+2\alpha^2\mathcal{K}^4}{1+2\alpha\mathcal{K}^2}\hat\nabla_\nu\mathcal{K}\mathcal{P}^\nu_\alpha\nonumber\\&&-2\frac{1+2\alpha\mathcal{K}^2+2\alpha^2\mathcal{K}^4}{1+2\alpha\mathcal{K}^2}\hat\nabla^\mu\sigma_{\mu\nu}\mathcal{P}^\nu_\alpha+\mathcal{O}(1)
\end{eqnarray}
Also,
\begin{equation}
\hat\nabla^\mu\sigma_{\mu\nu}\mathcal{P}^\nu_\alpha=\frac{\hat\nabla^2u_\beta+\mathcal{K}(u\cdot K)_\beta}{2}\mathcal{P}^\beta_\alpha+\mathcal{O}(1)
\end{equation}
Combining the above two equations along with the leading order vector membrane equation we see that 
$$\hat\nabla^\mu T_{\mu\nu}\mathcal{P}^\nu_\alpha=\mathcal{O}(1).$$
hence, the membrane stress tensor is conserved to leading order provided the leading order membrane equations are satisfied. The overall normalisation of the stress tensor is fixed by matching the total energy on a static spherical membrane with the mass of a static spherical black hole of horizon radius same as the radius of the membrane. 
\subsubsection{The $\eta/s$ ratio}
We define the coefficient of shear tensor in the membrane world-volume stress tensor (with the above mentioned normalisation) as the ``shear viscosity coefficient" `$\eta$' of the membrane. The ratio of the shear viscosity to entropy density of the membrane for EGB gravity is then given by
\begin{equation}\label{etas}
\eta = \frac{1}{16\pi}\left(\frac{1+\alpha{\cal K}^2}{1+B}\right),\quad s = \frac{1}{4}\left(1+2\alpha{\cal K}^2\right), \quad \frac{\eta}{s} = \frac{1}{4\pi}\left[1-2\alpha{\cal K}^2\frac{(1+\alpha{\cal K}^2)}{(1+2\alpha{\cal K}^2)^2}\right]
\end{equation} 
The ratio matches with the $\eta/s$ ratio of the dual boundary conformal fluid for EGB gravity derived in  \cite{Brigante:2007nu} upto linear order in $\alpha$. Ideally we should be computing the boundary stress tensor sourced by the membranes in $AdS$ space (see \cite{Dandekar:2017aiv} for similar analysis in Einstein-Hilbert gravity). But this match at linear order in $\alpha$ is in similar spirit to the match of the corresponding ratio in \cite{Dandekar:2017aiv}. 
\subsection{Stationary solutions}
At the leading order in large $D$ there is no entropy production even non-perturbatively in $\beta$. But we borrow intuition from the membrane equations at subleading order for Einstein-Hilbert gravity derived in \cite{Dandekar:2016fvw} and for EGB gravity perturbatively in $\beta$ here to invoke the condition on stationarity as the absence of shear for the membrane world-volume velocity field. Since, the velocity field is both divergenceless (using membrane equation) and shear-less, it has to be proportional to a timelike Killing vector field $k^\mu$ present in the membrane world-volume. i.e.
$$u^\mu=\gamma k^\mu.$$
where $\gamma$ is the normalization factor. Using this form of velocity the scalar membrane equation is trivially satisfied as $$k\cdot \nabla\gamma=0\quad (\because k \quad \text{is killing vector}).$$
The vector membrane equation becomes
\begin{equation}
\left[-(1+B)\frac{\hat\nabla_\nu \gamma}{\gamma} +(1-B)\frac{\hat\nabla_\nu {\cal K}}{{\cal K}} \right]{\cal P}^\nu_\sigma = {\cal O}(1/D)
\end{equation}
Which can be messaged into a more suggestive form in the following manner
\begin{equation}
\begin{split}
&\implies \left[-\frac{\hat\nabla_\nu \gamma}{\gamma} +\frac{(1-B)}{(1+B)} \frac{\hat\nabla_\nu {\cal K}}{{\cal K}} \right]{\cal P}^\nu_\sigma = {\cal O}(1/D)
\\&\implies \left[-\frac{\hat\nabla_\nu \gamma}{\gamma} + \frac{\hat\nabla_\nu {\cal K}}{{\cal K}} -\frac{2B}{(1+B)} \frac{\hat\nabla_\nu {\cal K}}{{\cal K}} \right]{\cal P}^\nu_\sigma = {\cal O}(1/D) \\&\implies \left[-\hat\nabla_\nu \ln \gamma + \hat\nabla_\nu \ln {\cal K} + \hat\nabla_\nu \ln \left(\frac{1+\alpha{\cal K}^2}{1+2\alpha{\cal K}^2}\right)\right]{\cal P}^\nu_\sigma = {\cal O}(1/D)\\
&\implies \nabla_\nu\ln \left(\frac{\mathcal{K}}{\gamma}\frac{1+\alpha\mathcal{K}^2}{1+2\alpha\mathcal{K}^2}\right)\mathcal{P}^\nu_\sigma=\mathcal{O}(1/D)\\&\implies \frac{\mathcal{K}}{\gamma}\frac{1+\alpha\mathcal{K}^2}{1+2\alpha\mathcal{K}^2}=\mathcal{C}\quad\text{where,}\quad \mathcal{C}\quad \text{is a constant}
\end{split}
\end{equation}
The constant can be fixed by comparing what the left hand side above evaluates for a static black hole in the leading large $D$ limit. Comparing with the temperature of static black hole for EGB gravity obtained in \cite{Myers:1988ze}, the constant can be fixed to be given by 
\begin{equation}\label{seq}
\frac{{\cal K}}{\gamma} \left(\frac{1+\alpha{\cal K}^2}{1+2\alpha {\cal K}^2}\right) = 4\pi T_0
\end{equation}
\subsection{Thermodynamics of the Static Black Hole}
The membrane stress tensor can be used to compute the energy of a spherical static membrane of radius $r_0$. This is given by
\begin{equation}
M = \Omega_{D-2}r_0^{D-2}T_{00} = \frac{(D-2)}{16\pi r_0} \left(1+\frac{\beta}{r_0^2}\right)r_0^{D-2} \Omega_{D-2}
\end{equation}
This exactly matches with the mass of static black holes in EGB gravity given in Eq $7$ of \cite{Myers:1988ze} with our ansatz metric and Eq $8$ in that paper then gives the right match of mass of static black hole with the result derived above.
Similarly, the leading order entropy current derived above can be used to compute the entropy of the corresponding static black hole to be given by
\begin{equation}
S_{ent} = \frac{r_0^{D-2}\Omega_{D-2}}{4}\left(1+2\frac{\beta}{r_0^2}\right)
\end{equation}
which matches at the leading order in $1/D$ with the corresponding expression in Eq $13$ of \cite{Myers:1988ze}. 

\subsection{Quasinormal mode frequencies for the static spherical black holes}
The effective membrane equations for Einstein-Hilbert gravity could successfully reproduce the spectrum of light quasi-normal modes about Schwarzschild black holes in the $D\rightarrow\infty$ limit (see\cite{Bhattacharyya:2015dva, Dandekar:2016fvw}). In this section we use the effective membrane equation to leading order (non-perturbative in $\beta$) to predict the light quasi-normal modes of the static spherical black hole of unit horizon size in EGB gravity. 
Consider background flat spacetime in the spherical polar coordinates
\begin{equation}
ds^2 = -dt^2+dr^2+r^2 d\Omega_{D-2}^2
\end{equation}
The membrane configuration which computes the light QNMs of static black holes is given by
\begin{equation}
r=1+\epsilon \delta r(t,a)\quad,\quad u=-dt+\epsilon \delta u_t(t,a) dt +\epsilon \delta u_a(t,a) d\theta^a  
\end{equation}
where, $\theta^a$ denotes the angular coordinates on the $D-2$ dimensional unit sphere. We work upto linear order in the strength of the fluctuation denoted by $\epsilon$.
The membrane equations upto linear order in $\epsilon$ for this configuration is given by
\begin{equation}\label{lve}
\left(1+\frac{\bar\nabla^2}{D}\right)\delta u_a + (1-\mathcal{B}) \bar\nabla_a \left(1+\frac{\bar\nabla^2}{D}\right)\delta r - \partial_t \partial_a \delta r-(1+\mathcal{B})\partial_t\delta u_a = 0
\end{equation}
 and the scalar membrane equation becomes
 \begin{equation}\label{lse}
 \bar\nabla_a\delta u^a  + D \partial_t \delta r = 0
 \end{equation}
 where the covariant derivative w.r.t. the metric on the unit sphere is denoted by $\bar\nabla$. And 
 \begin{equation}
 \mathcal{B} = \frac{\beta}{1+2\beta+2\beta^2}.
 \end{equation}
The fluctuations can be decomposed into the basis of scalar and vector spherical harmonics and the Fourier basis in time. This decomposition converts the membrane equations into a set of algebraic equations for each angular momentum number. The frequencies of scalar and vector modes are obtained by solving these algebraic equations and are given by
\begin{equation}
\omega_v = \frac{-i(l-1)}{1+\mathcal{B}}\quad,\quad \omega_s = \frac{-i(l-1)\pm \sqrt{(l-1)(1-\mathcal{B}^2 l)}}{1+\mathcal{B}}
\end{equation} 
The spectrum of light QNMs for scalar and vector modes were obtained from the effective equations in the mass-momentum formalism in \cite{Chen:2018vbv} (see section 5.1). Our results for the scalar and vector spectra match with that analysis. 
\subsubsection{Linear stability analysis of static spherical black holes}
From the above expressions of the light QNMs it is clear that the static black hole is unstable (i.e. the linearised modes grow in time) if
\begin{eqnarray}
&&1+\mathcal{B}\le 0\nonumber\\
i.e. && -1\le \beta\le-\frac{1}{2}
\end{eqnarray}
Interestingly, the leading order in large $D$ contribution to Wald entropy of the static black hole is negative for $\alpha\mathcal{R}\le-\frac{1}{2}$ and hence falls into the region of instability mentioned above. Hence, the value of the GB parameter is constrained by the requirement of linear stability and positivity of Wald entropy. 

Again looking at the scalar modes we see that for $l\ge \frac{1}{\mathcal{B}^2}$ the scalar mode is purely imaginary. If we confine ourselves to $\beta\ge-\frac{1}{2}$ then 
\begin{equation}
\frac{1}{\mathcal{B}^2}\ge1 \quad \text{ and hence for}\quad  l\ge \frac{1}{\mathcal{B}^2}\ge 1,
\end{equation}
we get purely imaginary scalar modes. Also, one of the scalar modes with frequency $\frac{-i(l-1)- \sqrt{(l-1)(1-\mathcal{B}^2 l)}}{1+\mathcal{B}}$ becomes unstable if 
\begin{eqnarray}
&&(l-1)-\sqrt{(l-1)(\mathcal{B}^2l-1)}\le 0\nonumber\\
&&\implies l-1\le \mathcal{B}^2l-1\nonumber\\
&&\implies \mathcal{B}^2\ge1
\end{eqnarray}
Which is not possible once we have $\beta\ge-\frac{1}{2}$. So to conclude, in the allowed regime of $\beta\ge-\frac{1}{2}$ all the linearised modes are stable. 
\section{Summary and Future Directions}
In this paper we have derived the equations governing the membranes dual to dynamical black holes in the large $D$ limit in two different regimes
\begin{itemize}
	\item Perturbatively in $\beta$ (the GB parameter) upto $\mathcal{O}(\beta)$ but upto subleading order in $1/D$ and
	\item Non-perturbatively in $\beta$ but upto leading order in $1/D$
\end{itemize}
We have also constructed an membrane entropy current in both of these cases from the membrane equations which satisfy a local form of second law of thermodynamics in the corresponding regimes. What we mean by this is that when the membrane equations are obeyed, the divergence of Entropy current is always positive definite. We find a non-trivial but positive definite entropy production in the perturbative $\beta$ regime. We have not yet mapped the membrane entropy current to the corresponding quantities in the black hole picture except in the stationary configuration. We have shown that the entropy obtained by integrating the entropy current on the membrane world-volume in the stationary configuration matches the Wald entropy of the dual stationary black holes. 

For the non-perturbative in $\beta$ regime we have written down a world-volume stress tensor from two conditions, namely:
\begin{itemize}
	\item The stress tensor is conserved on shell upto relevant order in $1/D$
	\item The energy of a static spherical membrane computed from the stress tensor matches the mass of the static spherical black hole solution of EGB gravity with the same radius.
\end{itemize} 
We have also constructed the effective membrane equation for stationary membranes in this regime. 

We have seen that both in Einstein-Hilbert gravity and the perturbative in $\beta$ analysis here that non-trivial entropy production takes place only at first subleading order in $1/D$. Hence, we would like to work out the dual membrane equations non-perturbatively in $\beta$ but upto subleading order in $1/D$ for EGB gravity. Once, we have the membrane equations it will be interesting to see if  we can construct an entropy current which satisfies the second law.  

It will also be interesting to understand more about the general theory of membranes which are dual to black holes in large $D$ gravity, which are consistent with the presence of an entropy current which has a positive definite divergence\footnote{We would like to accomplish a task similar to \cite{Bhattacharya:2011tra}, where the authors derived a theory of first order in derivative superfluid dynamics.}. The idea will be to put constraints on the transport coefficients of the membrane from the validity of second law which can then be thought of as the possible constraints on the most general classical theory of gravity.

It will also be interesting to work out a more detailed map between geometrical quantities on the black hole and on the membrane world-volume. Some progress in this direction for Einstein-Hilbert gravity has been made in \cite{Bhattacharyya:2018iwt}.This map is necessary to understand what the membrane entropy current evaluates to in the black hole picture so that we have a better understanding of the quantity that satisfies second law in terms of the black hole variables. This is also necessary to match the results obtained by us to the corresponding results in a different regime of dynamics in \cite{Bhattacharjee:2015yaa, Wall:2015raa} and \cite{Bhattacharyya:2016xfs}. In this paper we have managed to work out this map in the stationary case upto relevant orders in $\beta$ and $1/D$.

\section*{Acknowledgement}
We would like to thank Shiraz Minwalla, Sayantani Bhattacharyya, Nilay Kundu, R. Loganayagam, Suvrat Raju for many useful discussions. We would like to thank IISER Thiruvananthapuram and TIFR Mumbai for their hospitality during the initial stages of this project. A.S is supported by the Ambizione grant no. $PZ00P2_174225/1$ of the Swiss National Science Foundation (SNSF) and partially by the NCCR grant no. $51NF40-141869$ ``The Mathematics of Physics" (SwissMap). 

 \section*{Appendix}
 \appendix
 
 
 \section{The sources for the second order EGB equations}\label{detailedsources}
 In this appendix we write down the explicit forms of the sources for the EGB equations at second order in $1/D$. In all the expressions given below we will be using two different types of projectors given by
 \begin{eqnarray}
 &&\mathcal{P}_{MN}=\eta_{MN}-n_M n_N+u_M u_N\nonumber\\
 &&\Pi_{MN}=\eta_{MN}-n_M n_N
 \end{eqnarray}
 
 \subsection{The tensor sector}
 \begin{equation}
 \begin{split}
 &\mathcal{S}^{(0,T)}_{MN} = e^{-R}\left[\frac{{\cal K}}{D}(K_{AB}-\nabla_{(A}u_{B)})-(K_{CA}-\nabla_Cu_A)(K_{DB}-\nabla_Du_B){\cal P}^{CD}\right]{\cal P}^A_M {\cal P}^B_N
 \\&\text{and,}\\& \mathcal{S}^{(1,T)}_{MN}=\frac{{\cal K}^2}{D^2}e^{-2R}\Bigg[(1+e^R)K_{AC}K_{BD}{\cal P}^{CD}-2\nabla_{(A}u_{C}{\cal P}^{CD}K_{DB)} + (1-e^R)\nabla_{(A}u_{C)}\nabla_{(B}u_{D)}{\cal P}^{CD} \\& + (3+e^R)\nabla_{[A}u_{C]}\nabla_{[D}u_{B]}{\cal P}^{CD}- 2(1+e^R)\nabla_{((A}u_{C)}{\cal P}^{CD}\nabla_{[D}u_{B])}\\& +\left(\frac{{\cal K}}{2D}\left(-4+2(-2+e^R)R+R^2\right)+u\cdot K\cdot u\right)K_{AB}\\& + \left(-\frac{{\cal K}}{2D}\left(-4-4R+R^2+2e^R(1+R)\right)+u\cdot K\cdot u\right)\nabla_{(A}u_{B)}\\& -2e^R (u\cdot\nabla u_A-u^CK_{CA})(u\cdot\nabla u_B-u^DK_{DB})\\& -\frac{{\cal K}}{D}u\cdot K\cdot u+\frac{{\cal K}^2}{2D^2}\left(-2-4R+R^2+2e^R(1+R)\right){\cal P}_{AB}\Bigg]{\cal P}^A_M {\cal P}^B_N
 \end{split}
 \end{equation}
 \subsection{The vector sector}
 \begin{equation}
 \begin{split}
 &S^{(0,V)}_M = -Re^{-R}\frac{D}{{\cal K}}\left[\frac{{\cal K}^2}{D^2}n^A \nabla_{(A}u_{F)}+\frac{{\cal K}}{D}u^D K_{DE}(K_{AF}-\nabla_Au_F){\cal P}^{EA}-\frac{\nabla_D {\cal K}}{D}{\cal P}^{CD}(K_{CF}-\nabla_C u_F)\right]{\cal P}^F_M \\&\text{and,}\\&S^{(1,V)}_M= \frac{{\cal K}^2}{D^2}e^{-2R}\Bigg[\left( \frac{1-e^R}{2}\right)u^C \nabla_C(\Pi^A_E\nabla_A u_B\Pi^B_F)u^E -3(1-e^R)(1-R)\frac{\nabla^2 u_C}{\cal K}{\cal P}^{CB}\nabla_{(B}u_{F)} \\&+ (1-e^R)(-3+5R)\frac{\nabla^2 u_C}{\cal K}{\cal P}^{CB}\nabla_{[B}u_{F]} + (3-3R+e^R(-3+2R))\frac{\nabla^2 u_C}{\cal K}{\cal P}^{CB}K_{BF} \\& -(5-3R+e^R(-5+2R))u\cdot \nabla u_C{\cal P}^{CB} K_{BF} - (1-e^R)(-5+3R)u\cdot \nabla u_C {\cal P}^{CB} \nabla_{(B}u_{F)} \\& +5(1-e^R)(1-R)u\cdot \nabla u_C {\cal P}^{CB}  \nabla_{[B}u_{F]} + \frac{1-e^R(1+2R)}{2}u_A{K^A}_C{\cal P}^{CB}K_{BF}  \\& + Re^R u_A{K^A}_C{\cal P}^{CB}\nabla_{[B}u_{F]} + \frac{3-e^R(3+4R)}{2}u\cdot K\cdot u~ u^BK_{BF} + (-1+e^R(1+2R))\frac{u\cdot \nabla {\cal K}}{{\cal K}}u^BK_{BF} \\& -\frac{{\cal K}}{4D}R(-3(-4+R)+e^R(-10+3R))u^BK_{BF} + (-3+e^R(3+2R))u\cdot K\cdot u~u\cdot \nabla u_F \\& +2(1-e^R(1+R))\frac{u\cdot\nabla {\cal K}}{{\cal K}}~u\cdot \nabla u_F + \frac{{\cal K}}{2D}(4+6R-3R^2+e^R(-4-4R+3R^2))u\cdot \nabla u_F \\& +(1-e^R)u\cdot K\cdot u \frac{\nabla^2 u_F}{{\cal K}} + (-1+e^R) \frac{u\cdot\nabla {\cal K}}{{\cal K}}\frac{\nabla^2 u_F}{{\cal K}} + Re^R u_A{K^A}_C{\cal P}^{CB}\nabla_{(B}u_{F)}\\& +\frac{{\cal K}}{4D}(-8+3R^2-e^R(-8+2R+3R^2))\frac{\nabla^2 u_F}{{\cal K}} \Bigg]{\cal P}^F_M
 \end{split}
 \end{equation}
 \subsection{The scalar sector}
 \subsubsection{The trace sector}
 \begin{equation}
 \begin{split}
& \mathcal{S}^{(0,Tr)}=0\\&\text{and,}\\&\mathcal{S}^{(1,Tr)}=- 2\frac{{\cal K}^2}{D^2} e^{-R}\Bigg[(K_{CA}-\nabla_Cu_A)(K_{DB}-\nabla_Du_B){\cal P}^{AB}{\cal P}^{CD}-\frac{{\cal K}^2}{D-3}-2\frac{{\cal K}}{D}u\cdot K\cdot u+\frac{{\cal K}^2}{D^2}\Bigg]
 \end{split}
 \end{equation}
 \subsubsection{The $H^{(S)}$ sector}
 \begin{equation}
 \begin{split}
 &S^{(0,S)} =  e^{-R}\Bigg[ \nabla_A u_B \nabla_C u_D \Pi^{BC}\Pi^{AD} - 2 \nabla_{[A} u_{B]} \nabla_{[C} u_{D]} \Pi^{BC}\Pi^{AD} -\frac{R^2}{2}u\cdot \nabla u_C {\cal P}^{CA} u\cdot \nabla u_A \\&+2 u^BK_{BA}u\cdot \nabla u_C {\cal P}^{CA} +R^2u\cdot \nabla u_C{\cal P}^{CA}\frac{\nabla^2u_A}{\cal K}-2u_B{K^B}_C{\cal P}^{CA}\frac{\nabla^2u_A}{\cal K}+\frac{2-R^2}{2}\frac{\nabla^2u_A}{\cal K}\frac{\nabla^2u_C}{\cal K}{\cal P}^{CA}\\&-2 \left(\frac{u\cdot\nabla{\cal K}}{\cal K}-u\cdot K\cdot u\right)^2+\frac{-2+R^2}{2D}({\cal K}u\cdot K \cdot u-u\cdot \nabla{\cal K})\Bigg]
 \\&\text{and,}\\&S^{(1,S)} = \frac{{\cal K}^2}{D^2}e^{-R}\Bigg[ \frac{1}{2}e^{-R}(-8+3e^R(-1+R))\nabla^A \left(\frac{\hat\nabla^2\hat\nabla^2u_A}{{\cal K}^3}\right)+(-1+R)\nabla^A \left(\frac{\hat\nabla_A\hat\nabla^2{\cal K}}{{\cal K}^3}\right)\\&+K_{AB}K^{AB}-\frac{{\cal K}^2}{D-3}+(1-4e^{-R})\nabla_A u_B \nabla_C u_D {\cal P}^{AC}{\cal P}^{BD}-e^{-R}\nabla_{[A} u_{B]} \nabla_{[C} u_{D]} {\cal P}^{AC}{\cal P}^{BD} \\& -\frac{1}{2}e^{-R}(-4(2-R+R^2)+e^R(-10-4R+5R^2))\frac{\nabla^2u_A}{\cal K}\frac{\nabla^2u_B}{\cal K}{\cal P}^{AB}\\&-e^{-R}(-4+(2-4e^R)R+(-2+3e^R)R^2)u\cdot \nabla u_A u\cdot \nabla u_B {\cal P}^{AB}\\& +\frac{1}{2}e^{-R}(-8+(2-R+R^2)+e^R(-10-12R+11R^2))u\cdot \nabla u_B \frac{\nabla^2u_A}{\cal K} {\cal P}^{AB}  \\&+ 2(1+2e^{-R})u^BK_{BA}u^CK_{CD}{\cal P}^{AD} -\frac{1}{2}e^{-R}(24+e^R(-6+R^2))u^BK_{BA} u\cdot \nabla u_C{\cal P}^{AC} \\& + \frac{1}{2}e^{-R}(40+e^R(2-4R+R^2))\frac{\nabla^2u_C}{\cal K} u^BK_{BA}{\cal P}^{AC} +e^{-R}(-32+e^R(5-R+2R^2))u\cdot K\cdot u \frac{u\cdot \nabla {\cal K}}{{\cal K}} \\&-e^{-R}(-8+e^R(4+R^2))(u\cdot K \cdot u)^2 -e^{-R}(-36+R^2 e^R)\left(\frac{u\cdot \nabla {\cal K}}{\cal K}\right)^2 + \frac{1}{2}e^{-R}(-4(-5+R)R
 \\&+e^R(4-4R-4R^2+R^3))\frac{u\cdot \nabla {\cal K}}{D}  -\frac{1}{2}e^{-R}(-4(1-5R+R^2)+e^R(8-5R^2+R^3))\frac{{\cal K}}{D} u\cdot K\cdot u 
 \\& -\frac{1}{2}e^{-R}(e^R(-1+R)+2(-1-3R+R^2)) \frac{{\cal K}^2}{D^2}\Bigg]
 \end{split}
 \end{equation}
 where $\hat \nabla$ denotes spacetime derivative projected orthogonal to normal $n_M$. 
\bibliographystyle{JHEP}
\bibliography{ssbib}
\end{document}